\documentclass[ reprint, superscriptaddress,  amsmath,amssymb,  aps,  pre , showpacs , showkeys  ]{revtex4-1}
	\usepackage{sidecap}
	\sidecaptionvpos{figure}{c}
	\usepackage{graphicx}% Include figure files
	\usepackage{dcolumn}% Align table columns on decimal point
	\usepackage{bm}% bold math
	\usepackage[%
		colorlinks=true,
		linkcolor=blue,
		citecolor=blue,
		filecolor=blue,
		urlcolor=blue]{hyperref}
	\usepackage{amsmath,amsfonts, amsgen,amsbsy,amsopn,amstext,amssymb}
	\usepackage{epstopdf}
	\usepackage{color}
	\newcommand{\I}{\textrm{i}}
	\newcommand{\xx}{\mathbf{x}}
	\renewcommand{\(}{\left(}
	\renewcommand{\)}{\right)}
	
	\newcommand{\e}{\mathbf{e}}

\begin{document}
\preprint{}
%Title
	\title{General method to retrieve all effective acoustic properties of fully-anisotropic fluid materials in three dimensional space}
	%General acoustic characterization of three-dimensional homogeneous anisotropic fluid layers
%%%%%%%%%%%%%%%%%%%%%%%%%%%%%%%%%%%%%%%%%%%%%%%%%%%%%%%%%%%%%%%%%%
%Authors and affiliation
	\author{Arthur~Terroir}
		%\email{Arthur.Terroir.Etu@univ-lemans.fr}
	\author{Logan~Schwan}
	%\email{Logan.Schwan@univ-lemans.fr}
        \author{Th\'eo~Cavalieri}
	%\email{Theo.Cavalieri@univ-lemans.fr}
	\author{Vicente~Romero-Garc\'ia}
		%\email{Vicente.Romero@univ-lemans.fr}
	\author{Gw\'ena\"el~Gabard}
		%\email{Gwenael.Gabard@univ-lemans.fr}
	\author{Jean-Philippe~Groby}
		\email{Jean-Philippe.Groby@univ-lemans.fr}
	\affiliation{Laboratoire d'Acoustique de l'Universit\'e du Mans, LAUM - UMR CNRS 6613, Le Mans Universit\'e, Avenue Olivier Messiaen, 72085 LE MANS CEDEX 9, France}
%Date
	\date{\today}
%%%%%%%%%%%%%%%%%%%%%%%%%%%%%%%%%%%%%%%%%%%%%%%%%%%%%%%%%%%%%%%%%%
\begin{abstract}
Anisotropic fluid materials are of growing interest with the development of metamaterials and transformation acoustics. In the general three-dimensional case, such materials are characterized by a bulk modulus and a full symmetric matrix of density. Here, a method is presented to retrieve the bulk modulus and all six components of the density matrix from a selected set of six incident plane waves impinging on a layer of the material. From the six components of the density tensor, the three principal directions and the three principal densities of the material are recovered. The approach relies on the analytical expression of the reflection and transmission coefficients derived from a state vector analysis. It results in simple, closed-form, and easily-implementable  inverse relations for the material parameters. As an illustration, the case of sound propagation through an orthorhombic lattice of overlapping air-filled ellipsoids is considered, the effective complex and frequency-dependent bulk modulus and density matrix of which are derived from homogenization cell problems and account for viscothermal losses. The retrieval method is then applied to the homogenized layer and results bear testament to its robustness to extract accurately all seven material parameters. This makes possible the characterization and design of anisotropic fluid materials in three dimensions.
\end{abstract}
%%%%%%%%%%%%%%%%%%%%%%%%%%%%%%%%%%%%%%%%%%%%%%%%%%%%%%%%%%%%%%%%%%
% PACS, the Physics and Astronomy Classification Scheme.
%	\pacs{Valid PACS appear here}
%Keywords
%	\keywords{Anisotropic fluid; }
%Make title page
	\maketitle
%%%%%%%%%%%%%%%%%%%%%%%%%%%%%%%%%%%%%%%%%%%%%%%%%%%%%%%%%%%%%%%%%%

%%%%%%%%%%%%%%%%%%%%%%%%%%%%%%%%%%%%%%%%%%%%%%%%%%%%%%%%%%%%%%%%%%
%%%%%%%%%%%%%%%%%%%%%%%%%%%%%%%%%%%%%%%%%%%%%%%%%%%%%%%%%%%%%%%%%%
\section{Introduction}

With the rapid development of acoustic metamaterials  and transformation acoustics, efficient characterization methods that enable to estimate the unprecedented acoustic effective properties of structured materials are timely required. Characterization methods based on the inversion of the scattering matrix \cite{Nicolson1970,Weir1974} have been largely developed in the field of metamaterials \cite{Smith2002} and acoustic materials \cite{Song2000}. They are of particular interest in the design of acoustic metamaterials \cite{Popa2009,Zigoneanu2011,Jiang2011} since they directly provide their effective density and bulk modulus. Alternatively, these methods also turn out to be well-suited to retrieve the effective parameters of periodic arrangements of unit cells \cite{Fokin2007}, provided that their effective material supports only one propagative mode in the frequency range of interest and that Drude layers at  its boundaries are accounted for at high frequencies \cite{Simovski2007}. 

However, many acoustic metamaterials may be described as effective anisotropic  fluids \cite{Christensen2012,Torrent2008} notably to achieve acoustic cloaking \cite{Norris2015}. 
Certainly, characterization methods have been extended to characterize three-dimensional anisotropic materials with principal directions belonging to the layer plane interface \cite{Li2009,Jiang2011}, or two-dimensional anisotropic materials with principal directions arbitrarily tilted with respect to the reference coordinate system \cite{Castanie2014, Park2016}. Nevertheless, no specific methods seem to have been developed to characterize fully anisotropic acoustic materials in three dimensions (3-D).  Our aim here is to present a general retrieval method to extract the bulk modulus and all six components of the 3-D symmetric anisotropic tensor of density from a limited number of characterization tests. To do so, we build upon past work to extend methods based on plane wave reflection and transmission through a layer sample. Here, the general 3-D case of fully anisotropic fluid material having  principal axes tilted in a-priori unknown directions is considered.

The article is organized as follows. In Sec. \ref{sec:DirectProblem}, the direct problem is solved via a state vector formalism to yield the reflection and transmission coefficients. In particular, the transmission coefficient  is shown to exhibit phase delays which are related to the orientation of the material with respect to the layer interfaces. These phase delays will appear to be of paramount importance in the retrieval method. In Sec. \ref{sec:InverseProblem}, the inverse problem is studied and the general retrieval method is presented. It provides the analytical expression of the bulk modulus, and all six coefficients of the density tensor as functions of the reflection and transmission coefficients obtained from interrogation of the layer by incident plane waves at specific angles of incidence and orientation of the incident plane. In Sec. \ref{sec:Application}, the efficiency of the procedure is demonstrated in the case of  sound propagation through an anisotropic viscothermal fluid layer made of an orthorhombic lattice of overlapping ellipsoids. The effective poroacoustic properties of the array are first derived from the theory of two-scale asymptotic homogenization \cite{SanchezPalencia1980,Auriault2009} and the retrieval method is then applied to the homogenized anisotropic layer. All seven  material parameters (6 coefficients of the symmetric density tensor and bulk modulus) are accurately retrieved. They provide insight on the orientation of the material microstructure, through the recovery of the three principal directions and principal densities.

%%%%%%%%%%%%%%%%%%%%%%%%%%%%%%%%%%%%%%%%%%%%%%%%%%%%%%%%%%%%%%%%%%
%%%%%%%%%%%%%%%%%%%%%%%%%%%%%%%%%%%%%%%%%%%%%%%%%%%%%%%%%%%%%%%%%%

%%%%%%%%%%%%%%%%%%%%%%%%%%%%%%%%%%%%%%%%%%%%%%%%%%%%%%%%%%%%%%%%%%
%%%%%%%%%%%%%%%%%%%%%%%%%%%%%%%%%%%%%%%%%%%%%%%%%%%%%%%%%%%%%%%%%%
\section{Direct problem}
\label{sec:DirectProblem}

\begin{figure}[tbp]
	\centering
		\includegraphics[width=8.5cm]{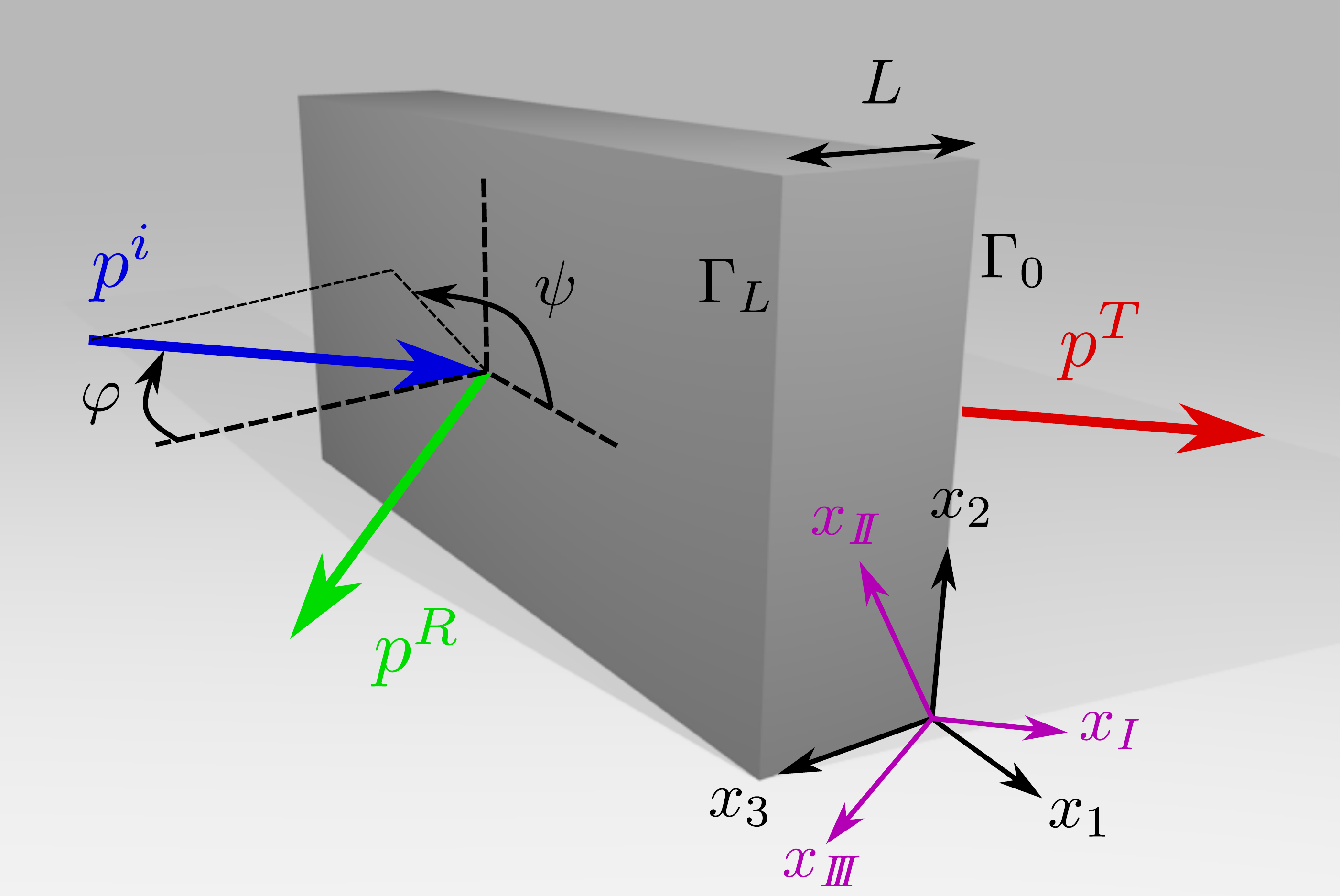}
	\caption{\label{fig:FIG1} (Color online) Conceptual view of the homogeneous anisotropic fluid layer of thickness $L$. The principal directions of the anisotropic fluid are denoted by $(x_{I},x_{I\!\!I},x_{I\!\!I\!\!I})$.}
\end{figure}

In this section, the plane wave propagation through a layer made of homogeneous anisotropic fluid material $\Omega$ is studied, see Fig. \ref{fig:FIG1}. 
The layer has the thickness $L$ and its constitutive material $\Omega $ has the bulk modulus $B$ and density tensor $\boldsymbol{\rho}$. In the reference Cartesian coordinate system $\mathcal{R}_0=\left(O,\e_1,\e_2,\e_3 \right)$ with position coordinates $(x_1,x_2,x_3)$, the mutually parallel plane boundaries $\Gamma_0$ and $\Gamma_L$ of the layer are given by the equations $x_3=0$ and $x_3=L$ respectively. The layer is surrounded on both sides $x_3\leq 0$ and $x_3\geq L$ by a homogeneous isotropic fluid $\Omega_0$ of scalar density $\rho_0$ and bulk modulus $B_0$. It leads to the sound speed  $c_0=\sqrt{B_0/\rho_0}$ and characteristic impedance $Z_0=\rho_0 c_0$ in the surrounding medium $\Omega_0$. Here, the analysis is performed in the linear harmonic regime at the circular frequency $\omega$ with the implicit time dependence $e^{-\I\omega t}$. In this system, the pressure and particle velocity fields $(P, \mathbf{V})$ in the layer and $(p, \mathbf{v})$ in the surrounding medium, are governed by the equations of mass and momentum conservation: 
%----------------------------------------------------------------------------------%
\begin{subequations}
	\begin{align}
		&& \I\omega P\big/ B= \nabla \cdot \mathbf{V}    \quad \text{and}\quad  \I  \omega \boldsymbol{\rho} \cdot \mathbf{V}=\nabla P \quad \text{in $\Omega$,}
		\label{eq:GoverningEqnsA}\\
		&& \I  \omega p\big/B_0= \nabla \cdot  \mathbf{v}   \quad \text{and}\quad  \I  \omega \rho_0 \mathbf{v}=\nabla  p \quad \text{in $\Omega_0$.}
		\label{eq:GoverningEqnsB}%
	\end{align}
\label{eq:GoverningEqns}%
\end{subequations}
%----------------------------------------------------------------------------------%
Equations (\ref{eq:GoverningEqnsA}) and (\ref{eq:GoverningEqnsB}) testify that the anisotropy of the material $\Omega$ in the layer is accounted for by the tensorial character of its density. 
As usual for passive media, the density tensor $\boldsymbol{\rho}$ is symmetric, that is  $\:^t\boldsymbol{\rho}=\boldsymbol{\rho}$ where $\:^t$ denotes transposition. 
In particular, the orthonormal coordinate system $\mathcal{R}_{\rho}=(\e_{I},\e_{I\!\!I},\e_{I\!\!I\!\!I})$ of its principal directions with coordinates  $(x_{I},x_{I\!\!I},x_{I\!\!I\!\!I})$  can be defined so that the density matrix is diagonal in this system. In other words, the density tensor can be written as $\boldsymbol{\rho} = \boldsymbol{\rho}^{\star}=\textbf{diag}\left(\rho_I,\rho_{I\!\!I},\rho_{I\!\!I\!\!I} \right)$ in $\mathcal{R}_{\rho}$, 
where $\rho_I$, $\rho_{I\!\!I}$ and $\rho_{I\!\!I\!\!I} $ are the principal densities.  As a result, when expressed in the reference coordinate system $\mathcal{R}_0$, the density tensor reads $\boldsymbol{\rho}= \mathbf{R}\cdot  \boldsymbol{\rho}^{\star} \cdot \,^t\mathbf{R}$ where $\mathbf{R}=\mathbf{R}_3\left(\theta_{I\!\!I\!\!I} \right)\mathbf{R}_2\left(\theta_{I\!\!I} \right) \mathbf{R}_1\left(\theta_{I} \right)$ is the rotation matrix between the two coordinate systems, with $\mathbf{R}_1$, $\mathbf{R}_2$, $\mathbf{R}_3$ being elementary matrices of rotations  and $\theta_I$, $\theta_{I\!\!I}$ and $\theta_{I\!\!I\!\!I}$ the  roll, pitch, and yaw angles. Moreover, it is worth recalling that, as effective properties, the bulk modulus $B$ and density tensor $ \boldsymbol{\rho}$ can be complex-valued and frequency-dependent. 

%***************************************************************%
%***************************************************************%

Now, the layer is submitted to the incident plane wave $p^i=e^{\I  k_1 x_1+\I  k_2 x_2-\I  k_3 \left(x_3-L\right)}$ propagating with a unit amplitude in the domain $x_3\geq L$ with the wavenumbers 
%----------------------------------------------------------------------------------%
\begin{subequations}
	\begin{align}
		&k_1=-k_0 \sin{\varphi}\cos{\psi}, 
		\label{eq:IncidentWavenumbersA}\\
		&k_2=-k_0 \sin{\varphi}\sin{\psi}, 
		\label{eq:IncidentWavenumbersB}\\
		&k_3=\sqrt{k_0^2-k_1^2-k_2^2}=k_0 \cos{\varphi}, 
		\label{eq:IncidentWavenumbersC}
	\end{align}
\label{eq:IncidentWavenumbers}%
\end{subequations}
%----------------------------------------------------------------------------------%
where $k_0=\omega/c_0$ is the acoustic wavenumber in $\Omega_0$, while  $\psi$ and $\varphi$ are the azimuthal and elevation angles measured from $(O,x_1)$ and $(O,x_3)$ respectively. 
Due to the  linearity in the system,  the Snell-Descartes Law holds: the in-plane wavevector $ \textbf{k}_\Gamma = k_1\e_1+k_2\e_2$ of the incident field is prescribed to the fields in both $\Omega_0$ and $\Omega$. In the surrounding medium $\Omega_0$, this gives rise to the specularly reflected  and transmitted waves $p^R$ and $p^T$ in the form
%----------------------------------------------------------------------------------%
\begin{equation}
	p^R=R e^{\I  k_3 \left(x_3-L\right)}  e^{\I  \textbf{k}_\Gamma  \cdot \xx_\Gamma }, 
\qquad 
	p^T=T e^{-\I  k_3 x_3} e^{\I \textbf{k}_\Gamma  \cdot \xx_\Gamma },
\label{eq:ReflectedTransmitted}
\end{equation}
%----------------------------------------------------------------------------------%
in domains $x_3\geq L$ and   $x_3\leq 0$ respectively, where $R$ and $T$ are the pressure reflection and transmission coefficients, while $\xx_\Gamma = x_1\e_1+x_2\e_2$ is the in-plane position vector. 
 In the layer  $\Omega$, the Snell-Descartes Law implies that the pressure and velocity fields take the form 
%----------------------------------------------------------------------------------%
\begin{equation}
	P=\widehat{P}(x_3)e^{\I   \textbf{k}_\Gamma  \cdot \xx_\Gamma }
\quad \text{and} \quad 
	\mathbf{V}=\widehat{\mathbf{V}}(x_3) e^{\I  \textbf{k}_\Gamma  \cdot \xx_\Gamma }, 
\label{eq:FieldsInLayer}
\end{equation}
%----------------------------------------------------------------------------------%
where $\widehat{P}(x_3)$ and $\widehat{\mathbf{V}}(x_3) $ are independent of $\xx_\Gamma$ due to the homogeneity of the layer, but still depend on the coordinate $x_3$. Substitution of Eq. (\ref{eq:FieldsInLayer}) into (\ref{eq:GoverningEqnsA})  leads to the following  equations of apparent mass and momentum conservation involving $\widehat{P}(x_3)$ and normal velocity $\widehat{V}_3=\e_3\cdot \widehat{\mathbf{V}}(x_3) $.
%----------------------------------------------------------------------------------%
\begin{subequations}
	\begin{align}
		&  \I   \omega \widehat{P} \big/ \widetilde{B} =    \I   (\textbf{q}\cdot \textbf{k}_\Gamma )  \widehat{V}_3  + \partial \widehat{V}_3 \big/\partial x_3  , 
		\label{eq:ApparentMassMomentumA}\\
		&   \I   \omega \widetilde{\rho} \,  \widehat{V}_3 =  \I   (\textbf{q}\cdot \textbf{k}_\Gamma )  \widehat{P} + \partial \widehat{P} \big/\partial x_3.
		\label{eq:ApparentMassMomentumB}
	\end{align}
\label{eq:ApparentMassMomentum}%
\end{subequations}
%----------------------------------------------------------------------------------%
Details about the derivation of Eqs.~(\ref{eq:ApparentMassMomentum}a,b) are provided in the Supplementary Material. 
In these equations,  the dimensionless vector  $\textbf{q}= q_1\e_1+q_2\e_2$ is induced by the anisotropic material which couples in-plane and normal directions, while scalars  $\widetilde{B}$ and $\widetilde{\rho}\, $ are the apparent bulk modulus and density. Denoting the inverse of the density tensor by  the symmetric tensor $\boldsymbol{H}=\boldsymbol{\rho}^{-1}$, the coefficients $q_1$, $q_2$, and the apparent density $ \widetilde{\rho}\,  $ are found to depend only on the (inverse) density tensor:
%----------------------------------------------------------------------------------%
\begin{equation}
	q_1 =    H_{13}\big/H_{33},\quad
	q_2 =    H_{23}\big/H_{33},\quad
	 \widetilde{\rho}\,  =1\big/H_{33} , 
\label{eq:ApparentParameters}
\end{equation}
%----------------------------------------------------------------------------------%
while the apparent bulk modulus $ \widetilde{B}  $ depends on the (inverse) density tensor, the bulk modulus $B$ and, more importantly, on the in-plane wavevector $\textbf{k}_\Gamma$ according to
%----------------------------------------------------------------------------------%
\begin{subequations}
	\begin{align}
		&\frac{\omega^2}{B} -\frac{\omega^2}{ \widetilde{B}(k_1,k_2)  }   = \xi_{11}  k_1^2 +  \xi_{22}k_2^2   +2 \,  \xi_{12} k_1k_2 , 
		\label{eq:ApparentBulk2A}\\
		&\text{with}\quad  \xi_{ij} = H_{ij}  - H_{33} \,  q_i \, q_j ,\ \  \forall (i,j)\in\{1,2\}^2. 
		\label{eq:ApparentBulk2B}%
	\end{align}
\label{eq:ApparentBulk}%
\end{subequations}
%----------------------------------------------------------------------------------%
It is worth noting that apparent density $\widetilde{\rho}_0=\rho_0$  and bulk modulus  $ \widetilde{B}_0 = B_0 k_0^2 \big/[k_0^2- k_1^2-k_2^2] $
in the isotropic surrounding medium $\Omega_0$ also display similar features,  but  the coupling vector $\textbf{q}$ and the coefficient $\xi_{12}$ are zero in $\Omega_0$. 

Introducing now the state vector $\boldsymbol{W}=\,^t\lbrace \widehat{P},\,\widehat{V}_3\rbrace$, the differential system in Eqs.~(\ref{eq:ApparentMassMomentum}a,b) can be cast in the following form, which is close to that of an homogeneous isotropic fluid, with the exception of non-zero diagonal terms induced by the anisotropic material: 
%----------------------------------------------------------------------------------%
\begin{equation}
  \frac{\partial \boldsymbol{W}}{\partial x_3} = \textbf{M}. \boldsymbol{W}
\quad \text{with}\quad  \textbf{M} =
\left[
    \begin{array}{cc}
      -\I  \textbf{q}\cdot \textbf{k}_\Gamma  & \I   \omega \widetilde{\rho}\, \\
     \displaystyle  \I   \omega \big/ \widetilde{B} & -\I  \textbf{q}\cdot \textbf{k}_\Gamma 
    \end{array}
    \right]. 
\label{eq:Sol1}
\end{equation}
%----------------------------------------------------------------------------------%
This system is solved by means of matrix  exponential, 
%----------------------------------------------------------------------------------%
\begin{equation}
	\boldsymbol{W}(L)=e^{\textbf{M}L}\cdot\boldsymbol{W}(0) .
\label{eq:Sol1b}
\end{equation}
%----------------------------------------------------------------------------------%
Due to the continuity of the pressure and normal component of the particle velocity at the layer boundaries  $\Gamma_0$ and $\Gamma_L$, the values of the state vector $\boldsymbol{W}$ at the layer boundaries $x_3=0$ and $x_3=L$ read
%----------------------------------------------------------------------------------%
\begin{equation}
	\boldsymbol{W}(L)= \begin{Bmatrix} 	R+1\\	 \\	\displaystyle	 \frac{R-1}{\widetilde{Z}_0} \\			\end{Bmatrix} 
\quad\text{and}\quad
	\boldsymbol{W}(0)=  \begin{Bmatrix} 	T \\ \\	\displaystyle			 \frac{-T }{\widetilde{Z}_0} \\			\end{Bmatrix}, 
\label{eq:ValueStateVectorBoundary}%
\end{equation}
%----------------------------------------------------------------------------------%
where  $\widetilde{Z}_0=(\widetilde{\rho}_0\widetilde{B}_0)^{1/2} = Z_0/\cos \varphi$ is the apparent impedance of air in the direction $(O,\e_3)$.
To calculate the exponential of the constitutive matrix $\textbf{M}$ in Eq.~(\ref{eq:Sol1b}), this latter  is  diagonalized according to
%----------------------------------------------------------------------------------%
\begin{equation}
	\textbf{M}~= \textbf{U}\cdot \boldsymbol{\Sigma} \cdot\textbf{U}^{-1} ,
\label{eq:ConstitutiveMatrix}
\end{equation}
%----------------------------------------------------------------------------------%
where $\boldsymbol{\Sigma}$ is the diagonal matrix  of eigenvalues and $\textbf{U}$ the matrix of associated eigenvectors: 
%----------------------------------------------------------------------------------%
\begin{subequations}
	\begin{align}
		&\boldsymbol{\Sigma}= 
			\begin{bmatrix} 
				\I  k_3^{-}& 0\\
				0 &\I  k_3^{+}\\
			\end{bmatrix} \quad \text{with}\quad k_3^{\pm} = -\textbf{q}\cdot \textbf{k}_\Gamma \pm\widetilde{k} ,
		\label{eq:SolA}\\
		&
 \textbf{U} = \frac{1}{\sqrt{2}}
	\begin{bmatrix}
		    \widetilde{Z}  & \widetilde{Z} \\
		    -1 & 1 \\
	\end{bmatrix}, \quad \text{with}\quad 
 \textbf{U}^{-1} = 
	\frac{1}{\sqrt{2}}
	\begin{bmatrix}
	     1\big/\widetilde{Z}  & -1\\
		1\big/\widetilde{Z}  & 1 \\
	\end{bmatrix}.
		\label{eq:SolB}%
	\end{align}
\label{eq:Sol}%
\end{subequations}
%----------------------------------------------------------------------------------%
Here, the wavenumber $\widetilde{k}$ and impedance $\widetilde{Z} $ are built from the apparent density $\widetilde{\rho}\, $ and bulk modulus $\widetilde{B}$ as
%----------------------------------------------------------------------------------%
\begin{equation}
	\widetilde{k}=  \omega\sqrt{\widetilde{\rho}\, \big/\widetilde{B}}
\qquad\text{and}\qquad
	\widetilde{Z} =\sqrt{\widetilde{\rho}\, \widetilde{B}}.
\label{eq:k3andZ3}
\end{equation}
%----------------------------------------------------------------------------------%
The eigenvalues  $k_3^{\pm}$  of the matrix $\textbf{M}$ in Eq.~(\ref{eq:SolA}) provide the dispersion relation in the anisotropic fluid. Indeed, the pressure field $\widehat{P}(x_3)$ will take the form $\widehat{P}=\widehat{P}^+ e^{ \I   k_3^{+}x_3} +\widehat{P}^- e^{ \I   k_3^{-}x_3}  $, where $\widehat{P}^{\pm}$ are complex amplitudes and $\widehat{P}^\pm e^{ \I   k_3^{\pm}x_3}$ represent waves propagating in the direction $\pm x_3$. Conversely to isotropic media, the wavenumbers $k_3^{\pm}$ are not necessarily opposite: their sum yields $k_3^{-}+k_3^{+}=2\textbf{q}\cdot \textbf{k}_\Gamma$, which will be  at the origin of phase delays in the transmission coefficients, as shown later. This effect is due to the coupling between the directions of the reference coordinate system operated by the anisotropic material when the density matrix is fully-symmetric. However, the coupling vector $\textbf{q}$ between in-plane and normal directions is zero when one principal direction of the (inverse) density tensor coincides with the direction $(O,x_3)$ normal to the boundaries of the layer. Then, the anisotropy of the material $\Omega$ only influences the apparent bulk modulus $\widetilde{B}$ according to Eq.~(\ref{eq:ApparentBulk}), which would depend only on the rotation of the principal directions around $(O,x_3)$. 

Finally, substitution of the boundary conditions  (\ref{eq:ValueStateVectorBoundary}) and of the diagonalized form (\ref{eq:ConstitutiveMatrix}) of the constitutive matrix $\textbf{M}$ into Eq.~(\ref{eq:Sol1b}),leads to the following linear system to solve for the reflection and transmission coefficients:
%----------------------------------------------------------------------------------%
\begin{equation}
	\begin{Bmatrix} 	R+1\\	 \\	\displaystyle	 \frac{R-1}{\widetilde{Z}_0} \\			\end{Bmatrix}
= \textbf{U} \cdot
	\begin{bmatrix} 	e^{ \I  k_3^{-}L} & 0\\  \\		0 & e^{\I  k_3^{+} L}\\		\end{bmatrix}
	\cdot 
	\textbf{U}^{-1}
	\cdot
	\begin{Bmatrix} 	T \\ \\	\displaystyle			 \frac{-T }{\widetilde{Z}_0} \\			\end{Bmatrix}.
\label{eq:Sol2}
\end{equation}
%----------------------------------------------------------------------------------%
Resolution of this linear system yields the following reflection and transmission coefficients:
%----------------------------------------------------------------------------------%
\begin{subequations}
\begin{align}
%....................................................................................................................%
	&R = \frac{ 		-\I  \left(\widetilde{Z}\big/\widetilde{Z}_0  - \widetilde{Z}_0\big/\widetilde{Z}    \right)  \sin (\widetilde{k}L )
			}{ 			2 \cos (\widetilde{k}L ) -\I  \left( \widetilde{Z}\big/\widetilde{Z}_0  + \widetilde{Z}_0\big/\widetilde{Z}     \right) \sin (\widetilde{k}L ) },  
\label{eq:CoeffR} \\
%....................................................................................................................%
	&T=\frac{	2\,  e^{\I  (\textbf{q}\cdot \textbf{k}_\Gamma ) L}
			}{				2 \cos(\widetilde{k}L) -\I  \left( \widetilde{Z}\big/\widetilde{Z}_0  + \widetilde{Z}_0\big/\widetilde{Z}     \right)  \sin(\widetilde{k}L) }, 
\label{eq:CoeffT}%
%....................................................................................................................%
\end{align}%
\label{eq:CoeffRandT}%
\end{subequations}%
%----------------------------------------------------------------------------------%
As mentioned previously, Eq.~(\ref{eq:CoeffT}) shows that the transmission coefficient $T$ is affected by the phase delay
$ \textbf{q}\cdot \textbf{k}_\Gamma  L $ due to the anisotropy of the material. This property will be of paramount importance when presenting the retrieval method in the next section. 

%%%%%%%%%%%%%%%%%%%%%%%%%%%%%%%%%%%%%%%%%%%%%%%%%%%%%%%%%%%%%%%%%%
%%%%%%%%%%%%%%%%%%%%%%%%%%%%%%%%%%%%%%%%%%%%%%%%%%%%%%%%%%%%%%%%%%

%%%%%%%%%%%%%%%%%%%%%%%%%%%%%%%%%%%%%%%%%%%%%%%%%%%%%%%%%%%%%%%%%%
%%%%%%%%%%%%%%%%%%%%%%%%%%%%%%%%%%%%%%%%%%%%%%%%%%%%%%%%%%%%%%%%%%
\section{Retrieval method}
\label{sec:InverseProblem}

The problem now consists in retrieving the six components of the symmetric density tensor, and the value of the bulk modulus from the knowledge of the thickness $L$ of the layer and the reflection and transmission coefficients, 
$R$ and $T$, at specific azimuthal and elevation angles $\psi$ and $\varphi$, that is at specific in-plane wavenumbers $\textbf{k}_\Gamma=(k_1,k_2)$, see Eq.~(\ref{eq:IncidentWavenumbers}). Once estimated, the material parameters will be marked by the superscript $^{\dag}$ in the form $(\rho_{11}^{\dag},\rho_{12}^{\dag},\rho_{13}^{\dag},\rho_{22}^{\dag},\rho_{23}^{\dag},\rho_{33}^{\dag}, B^{\dag})$.

Central to the retrieval method is the fact that the apparent impedance $\widetilde{Z} $ and wavenumber $\widetilde{k}$, and subsequently the apparent density $\widetilde{\rho}$ and bulk modulus $\widetilde{B}$,  can be directly retrieved from the reflection and transmission coefficients, assuming the prior knowledge of  the phase delay $ \textbf{q}\cdot \textbf{k}_\Gamma  L $ affecting the transmission coefficient in Eq.~(\ref{eq:CoeffT}). 
Indeed, inverting the system given by Eqs.~(\ref{eq:Sol2}) with the matrices expressed in Eqs.~(\ref{eq:Sol}a,b), the Nicolson--Ross--Weir procedure \cite{Nicolson1970,Weir1974} can be extended to oblique incidence and anisotropic media as follows, see Supplementary Material for details: 
%----------------------------------------------------------------------------------%
\begin{subequations}
\begin{align}
%....................................................................................................................%
&	\widetilde{Z}^\dag= \pm\widetilde{Z}_0
		 \sqrt{\frac{
			\left(Te^{-\I   \textbf{q}\cdot \textbf{k}_\Gamma L} \right)^2-\left(1+R \right)^2
		}{
			\left(Te^{-\I   \textbf{q}\cdot \textbf{k}_\Gamma L} \right)^2-\left(1-R \right)^2}} ,
\label{eq:CoeffZ}  \\
%....................................................................................................................%
& e^{\mp \I  \widetilde{k}^\dag L}=\, \chi^{\mp} =
 \(    1 + \displaystyle  \frac{ \widetilde{Z}_0  \mp  \widetilde{Z}^\dag }{\widetilde{Z}_0 \pm \widetilde{Z}^\dag}  R  \)
\frac{1}{ T e^{-\I   \textbf{q}\cdot \textbf{k}_\Gamma L}} .
  \label{eq:ExpKL}   
%....................................................................................................................%
\end{align}%
\label{eq:CoeffZandExpKL}%
\end{subequations}%
%----------------------------------------------------------------------------------%
While the sign in Eq.~(\ref{eq:CoeffZ}) is actually determined by the passivity constraint $\textrm{Re}(\widetilde{Z}^\dag )\geq0$, 
both signs in Eq.~(\ref{eq:ExpKL}) are physically sound. However, inverting $e^{- \I  \widetilde{k}^\dag L}$ is preferred here since  the negative $x_3$-going waves usually carry more energy than the positive ones for negative $x_3$-going incident wave. That provides 
%----------------------------------------------------------------------------------%
\begin{equation}
	\widetilde{k}^\dag= \left(-\textrm{ang}\left(\chi^- \right)+\I   \log |\chi^-|+2 n \pi \right) /L ,
\label{eq:Retrievedk3}
\end{equation}
%----------------------------------------------------------------------------------%
where $\textrm{ang}$ is the phase angle, $\log$ is the natural logarithm. In Eq.~(\ref{eq:Retrievedk3}), the term $2 n \pi$ with integer $n\in \mathbb{Z}$ is used to unwrap the phase of $\chi^-$ so that 
$\widetilde{k}^\dag$ is continuous over the frequencies, with $\widetilde{k}^\dag=0$ at the frequency $\omega=0$. The  integer $n$ has to be determined and depends on the nature of the material, but 
it is is usually zero when initiating the procedure at very low frequency. Further, using Eq.~(\ref{eq:k3andZ3}) and the values of $\widetilde{Z} ^\dag$ and $\widetilde{k}^\dag$, the apparent density and bulk modulus are retrieved as 
%----------------------------------------------------------------------------------%
\begin{equation}
	\widetilde{\rho}^\dag =  \widetilde{Z}^\dag  \widetilde{k}^\dag  \big/  \omega 
\qquad\text{and}\qquad
	\widetilde{B}^\dag =  \omega \, \widetilde{Z}^\dag \big/  \widetilde{k}^\dag.
\label{eq:Rho3andB3}
\end{equation}
%----------------------------------------------------------------------------------%
It is important here to emphasize that Eqs.~(\ref {eq:CoeffZ}) to (\ref{eq:Retrievedk3})  actually hold for any in-plane wavevector $\textbf{k}_\Gamma=(k_1,k_2)$, provided that the coefficients  $ q_1 $ and $q_2$ are known to calculate the phase delay $ \textbf{q}\cdot \textbf{k}_\Gamma  L $ in Eq.~(\ref{eq:CoeffT}). Since   $ q_1 $ and $q_2$   are independent of the wavevector $\textbf{k}_\Gamma=(k_1,k_2)$, see Eq.~(\ref{eq:ApparentParameters}),  they can be retrieved by using two pairs of transmission coefficients as follows. Choosing $T(k'_1,0)$, $T(-k'_1,0)$ and $T(0,k'_2)$, $T(0,-k'_2)$ with wavenumbers $k'_1\neq0$ and $k'_2\neq0$ ensures the equality of the denominators of $T(k'_1,0)$ and $T(-k'_1,0)$ on the one hand, and those of $T(0,k'_2)$ and $T(0,-k'_2)$ on the other hand, see Eq. (\ref{eq:CoeffT}). This can be explained by the fact that denominators of reflection and transmission coefficients actually represent the dispersion relation of the anisotropic layer modes. Using this property, the coefficients $q_1^\dag$ and $q_2^\dag$ are retrieved from the following relations: 
%----------------------------------------------------------------------------------%
\begin{equation}
	 e^{2\I   q_1^\dag k'_1 L} =\frac{T(k'_1,0)}{T(-k'_1,0)}; 
\qquad
	e^{2\I   q_2^\dag k'_2 L} =\frac{T(0,k'_2)}{T(0,-k'_2)}.
\label{eq:Retrievedqj}
\end{equation}
%----------------------------------------------------------------------------------%
At this stage, from the four transmission coefficients using the wavevectors $\textbf{k}_\Gamma=\pm k'_1 \e_1$ and $\textbf{k}_\Gamma=\pm k'_2 \e_2$ with any non-zero wavenumbers $k'_1\neq0$ and $k'_2\neq0$, the coefficients   $q_1^\dag$ and $q_2^\dag$  have been retrieved. Therefore, the apparent parameters $\widetilde{Z}^\dag$ and $\widetilde{k}^\dag$, and consequently $\widetilde{\rho}^\dag$ and $\widetilde{B}^\dag$ are not only known for  $\textbf{k}_\Gamma=\pm k'_1 \e_1$ and $\textbf{k}_\Gamma=\pm k'_2 \e_2$, but they can now be assessed for any incident wave, using Eqs. (\ref{eq:CoeffZ}-\ref{eq:Rho3andB3}). This central property is used in what follows to retrieve the bulk modulus $B$ and all six components of the density tensor $\boldsymbol{\rho}$. Please note that to obtain $\widetilde{B}^{\dag1}\equiv \widetilde{B}^{\dag} (k_1',0)$ and $\widetilde{B}^{\dag2}\equiv\widetilde{B}^{\dag} (0,k_2')$, being respectively equal to $\widetilde{B}^\dag (-k_1',0)$ and $\widetilde{B}^\dag (0,-k_2')$, only two additional reflection coefficients are needed, i.e. $R(k_1,0)$ and $R(0,k_2)$.

To gain access to the physical bulk modulus $B$, a fifth characterization test at normal incidence $(k_1,k_2)=(0,0)$ is here considered, which provides, according to Eq.~(\ref{eq:ApparentBulk2A}), 
%----------------------------------------------------------------------------------%
\begin{equation}
	B^{\dag} = \widetilde{B}^\dag (0,0).
\label{eq:RetrievedBulk}
\end{equation}
%----------------------------------------------------------------------------------%
Once the bulk modulus $B^{\dag}$ is determined, it is straightforward to retrieve the coefficients $\xi_{11}$ and $\xi_{22} $ from the four characterization tests already performed with the wavevectors $\textbf{k}_\Gamma=\pm k'_1 \e_1$ and $\textbf{k}_\Gamma=\pm k'_2 \e_2$. Indeed, using Eq.~(\ref{eq:ApparentBulk2A}), the following relation holds:
%----------------------------------------------------------------------------------%
\begin{equation}
\forall j \in \{1,2\},  \quad 
	\xi_{jj}^{\dag}=   \frac{\omega^2}{(k'_j)^2} \(  \frac{1}{B^{\dag}} -   \frac{1}{\widetilde{B}^{\dag j}}  \).
\label{eq:RetrievedXiJJ}
\end{equation}
%----------------------------------------------------------------------------------% 
To be in a position to retrieve all coefficients of the density tensor, a sixth and final characterization test with the wavevector $\textbf{k}_\Gamma=(k''_1,k''_2)$ such that $k''_1\neq 0$ and $k''_2\neq 0$ is considered. As it happens, with the knowledge of $B^{\dag} $, $\xi_{11}^{\dag}$ and $\xi_{22}^{\dag}$, equation ~(\ref{eq:ApparentBulk2A}) is solved for $\xi_{12}$ to  yield: 
%----------------------------------------------------------------------------------%
\begin{equation}
	 \xi^{\dag}_{12} = 	 
		\frac{\omega^2}{2   k''_1 k''_2}  \(  \frac{1}{B^{\dag}} - \frac{ 1}{\widetilde{B}^{\dag3}}   \) 
		-   \frac{\xi_{11}^{\dag} \, k''_1 }{2    k''_2}  - \frac{\xi_{22}^{\dag} \, k''_2 }{2   k''_1  },
\label{eq:RetrievedXi12}
\end{equation}
%----------------------------------------------------------------------------------%
with $\widetilde{B}^{\dag3}\equiv\widetilde{B}^{\dag}(k''_1,k''_2)$. The coefficients of the inverse density tensor $\boldsymbol{H}$ are retrieved from Eqs.~(\ref{eq:ApparentParameters}) and  (\ref{eq:ApparentBulk2B}) to  provide the following relations, where $i,\, j\in\{1,2\}$:
%----------------------------------------------------------------------------------%
\begin{equation}
	H_{33}^{\dag} = \frac{1}{\widetilde{\rho}^\dag},
\quad 
	H_{ij}^{\dag} =   \xi_{ij}^{\dag} +  H_{33}^{\dag}   q_i^\dag q_j^\dag  , 
\quad 
	 H_{i3}^{\dag} = q_i^\dag  H_{33}^{\dag}.
\label{eq:InverseDensity}%
\end{equation}
%----------------------------------------------------------------------------------%
Finally, the density tensor can be obtained by inverting $\boldsymbol{H}$ numerically, or by using the following expressions derived in detail in the Supplementary Material: 
 %----------------------------------------------------------------------------------%
\begin{subequations}
	\begin{align}
		%.................................................................................................................................................%
		& 		\rho_{11}^{\dag} =  \frac{ \xi_{22}^{\dag} }{\Delta_\xi^{\dag}} , 
		\qquad  \rho_{22}^{\dag} =  \frac{\xi_{11}^{\dag}  }{\Delta_\xi^{\dag} }  ,
		\qquad  \rho_{12}^{\dag} =  \frac{ -   \xi_{12}^{\dag} }{ \Delta_\xi^{\dag} } ,
		\label{eq:RetrievedRhoInPlane} \\
		%.................................................................................................................................................%
		& 	\rho_{13}^{\dag} = \frac{ \xi_{12}^{\dag} q_2^{\dag} - \xi_{22}^{\dag} q_1^{\dag}}{ \Delta_\xi^{\dag}} ,
		\qquad 
			\rho_{23}^{\dag} = \frac{ \xi_{12}^{\dag} q_1^{\dag} - \xi_{11}^{\dag} q_2^{\dag}}{\Delta_\xi^{\dag}} ,
		\label{eq:RetrievedRhoCoupling} \\
		%.................................................................................................................................................%
		&		\rho_{33}^{\dag} =   \widetilde{\rho}\, ^{\dag} + \frac{1}{\Delta_\xi^{\dag}}\Big(   \xi_{22}^{\dag} (q_1^{\dag})^2   +\xi_{11}^{\dag} (q_2^{\dag})^2     -2\, \xi_{12}^{\dag} q_1^{\dag} q_2^{\dag}   \Big)  ,
		\label{eq:RetrievedRho33} 
		%.................................................................................................................................................%
	\end{align}
\label{eq:RetrievedRho}%
\end{subequations}
%----------------------------------------------------------------------------------%
where $ \Delta_\xi^{\dag} = \xi_{11}^{\dag}   \xi_{22}^{\dag}  -  ( \xi_{12}^{\dag}  )^2$. This brings an end to the retrieval procedure. The seven rheological parameters characterizing the anisotropic fluid material in the layer (the bulk modulus and the six coefficients of the density tensor) have been retrieved from six transmission coefficients and four associated reflection coefficients. They derive from tests performed  at normal incidence $\varphi\equiv0$;  at oblique incidence ($\varphi\neq0$) with opposite pairs of in-plane wavevectors $\textbf{k}_\Gamma=(\pm k'_1,0)$ and $\textbf{k}_\Gamma=(0,\pm k'_2)$ oriented along the axes  $\e_1$ and $\e_2$ of  the reference coordinate system;  and at oblique incidence ($\varphi\neq0$) with in-plane wavevector  $\textbf{k}_\Gamma=(k''_1,k''_2)$   out of the  axes of  the reference coordinate system, that is  $k''_1 k''_2 \neq 0$. 

%***************************************************************%
\begin{figure*}
\includegraphics[width=\textwidth]{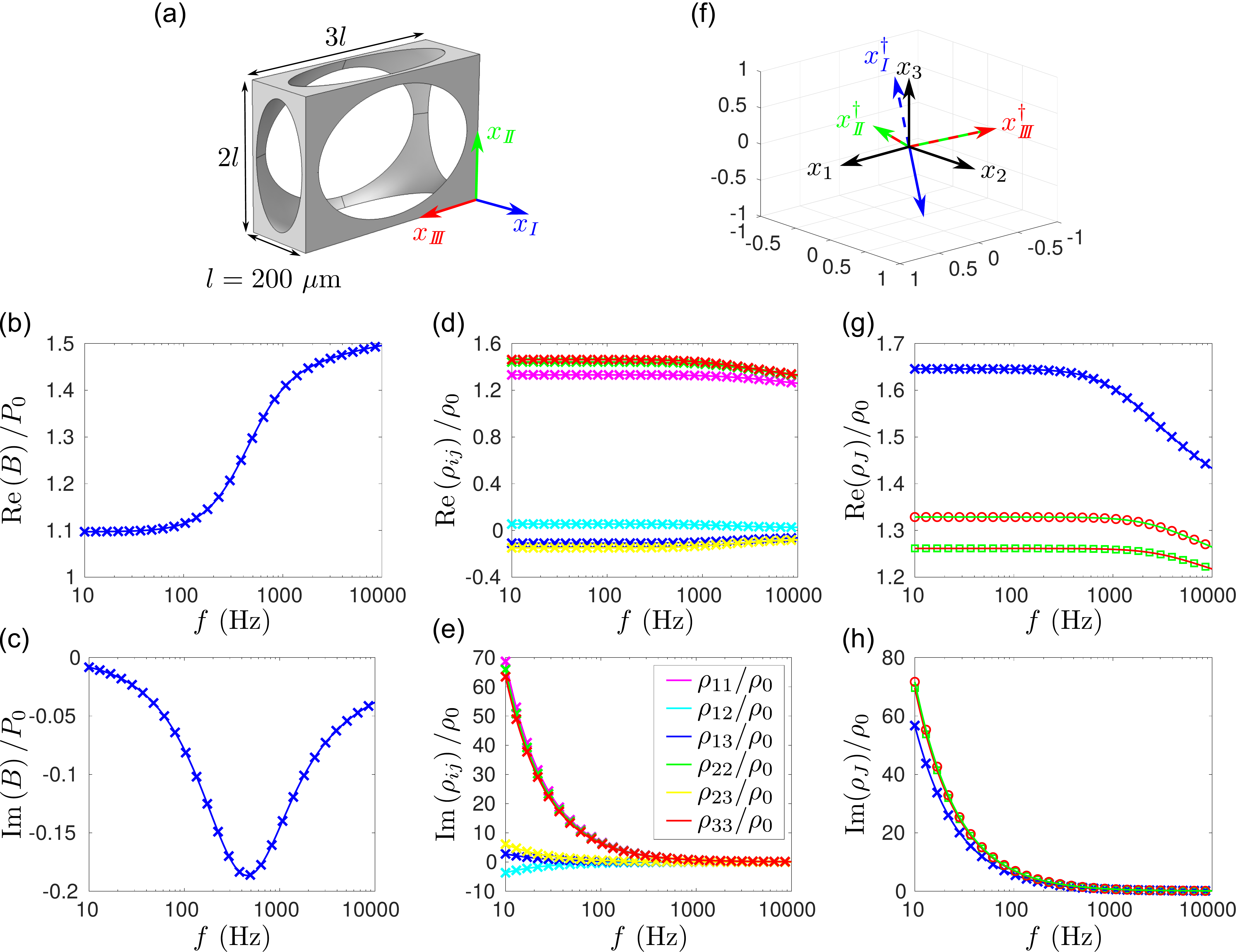}
\caption{\label{fig:FIG2}{(Color online) Schematic view of the unit cell (a). Reconstructed (${\color{blue}{\times}}$) and initial ($\color{blue}{\rule[0.5ex]{1em}{1pt}}$) real and imaginary parts of the normalized bulk (b-c). Reconstructed ($\times$) and initial ($\rule[0.5ex]{1em}{1pt}$) real and imaginary parts of the six normalized components of the symmetric density tensor (d-e). The inset Fig. (e) depicts the color of the different components. Initial and reconstructed principal directions (f): $x_I$ ($\color{blue}{\rule[0.5ex]{1em}{1pt}}$), $x_{I\!\!I}$ ($\color{green}{\rule[0.5ex]{1em}{1pt}}$), and $x_{I\!\!I\!\!I}$ ($\color{red}{\rule[0.5ex]{1em}{1pt}}$); $x_I^\dag$ ($\color{blue}{--}$), $x_{I\!\!I}^\dag$ ($\color{green}{--}$), and $x_{I\!\!I\!\!I}^\dag$ ($\color{red}{--}$). Initial and reconstructed real and imaginary parts of the normalized densities in the principal directions (g-h): $\rho_I$ ($\color{blue}{\rule[0.5ex]{1em}{1pt}}$), $\rho_{I\!\!I}$ ($\color{green}{\rule[0.5ex]{1em}{1pt}}$), and $\rho_{I\!\!I\!\!I}$ ($\color{red}{\rule[0.5ex]{1em}{1pt}}$); $\rho_I^\dag$ ($\color{blue}{\times}$), $\rho_{I\!\!I}^\dag$ ($\color{green}{\square}$), and $\rho_{I\!\!I\!\!I}^\dag$ ($\color{red}{\circ}$.)}}
\end{figure*}
%***************************************************************%

It can be interesting to note that the apparent density $ \widetilde{\rho}\,  $ (or equivalently $H_{33}=1/\widetilde{\rho}\,  $) can actually be estimated independently in the six tests with the prior knowledge of coefficients $q_1^{\dag}$ and/or $q_2^{\dag}$, see Eqs.~(\ref {eq:CoeffZ}) to (\ref{eq:Retrievedk3}). This property can be used to assess the accuracy of the retrieved parameters (all tests should provide the same  apparent density $ \widetilde{\rho}  $), and it can also lead  to estimate  $ \widetilde{\rho} $ precisely by averaging all its retrieved values. 

In the same line of thought, it is worth noting that the retrieval method has been presented on the basis of only six characterization tests. However, the wavenumbers $k'_1$ and $k'_2$ used for tests at oblique incidence in the directions $\e_1$ and $\e_2$, and  wavenumbers $k''_1$ and $k''_2$ used for tests at oblique incidence out of directions $\e_1$ and $\e_2$ are actually not specified. Several repetitions of these tests with various values of $k'_1$, $k'_2$, $k''_1$ and $k''_2$ (that is, various angles of incidence $\varphi$ for all tests and various azimuthal angles $\psi$ for the last test) can therefore be performed. This increases the number of tests but it would allow to average the various retrieved values of the material parameters, and hence smooth experimental or numerical noise in the initial data. 

Finally, once the full symmetric tensor of (inverse) density is retrieved, it can be diagonalized to yield $\boldsymbol{\rho}^\dag= \mathbf{R}^\dag\cdot  \boldsymbol{\rho}^{\star\dag} \cdot \:^t\mathbf{R}^\dag$ where $ \boldsymbol{\rho}^{\star\dag}$ is the diagonal matrix of the retrieved principal densities, see Sec. \ref{sec:DirectProblem}. In particular, the columns of the retrieved rotation tensor $\mathbf{R}^\dag$ actually represent the coordinates of the right-hand orthonormal eigenvectors of the density tensor in the reference coordinate system $\mathcal{R}_0$. This requires to normalize and orient correctly the eigenvectors. These latter provide precious insight on the orientation of the material axes with respect to  $\mathcal{R}_0$, which can be related to the orientation of the anisotropic layer microstructure.  It is worth recalling also that such orientation of material axes can be characterized by the roll, pitch, and yaw angles $\theta_I$, $\theta_{I\!\!I}$ and $\theta_{I\!\!I\!\!I}$. However, these Euler angles are not unique without additional constraints on the range of their values, and without a convention to specify in which order the principal densities are sorted in the diagonal density matrix $ \boldsymbol{\rho}^{\star\dag}$. Indeed, the order of  the principal densities in $ \boldsymbol{\rho}^{\star\dag}$  determines which eigenvector of the density tensor actually plays the role of vectors $\e_{I}$, $\e_{I\!\!I}$ or $\e_{I\!\!I\!\!I}$ in the material coordinate system $\mathcal{R}_\rho$.

%%%%%%%%%%%%%%%%%%%%%%%%%%%%%%%%%%%%%%%%%%%%%%%%%%%%%%%%%%%%%%%%%%
%%%%%%%%%%%%%%%%%%%%%%%%%%%%%%%%%%%%%%%%%%%%%%%%%%%%%%%%%%%%%%%%%%

%%%%%%%%%%%%%%%%%%%%%%%%%%%%%%%%%%%%%%%%%%%%%%%%%%%%%%%%%%%%%%%%%%
%%%%%%%%%%%%%%%%%%%%%%%%%%%%%%%%%%%%%%%%%%%%%%%%%%%%%%%%%%%%%%%%%%
\section{Application and validation}
\label{sec:Application}

In this section, the efficiency of the retrieval method to estimate accurately the bulk modulus and all six components of the density tensor is demonstrated. 
Here, the procedure is applied to an anisotropic viscothermal fluid layer of thickness $L=3\textrm{ cm}$ under ambient conditions. 
For numerical applications, the air density, adiabatic constant, dynamic viscosity, specific heat capacity at constant pressure, and thermal conductivity at equilibrium are 
 $\rho_0=1.213\textrm{ kg.m}^{-3}$, $\gamma=1.4$,   $\eta=1.839\times 10^{-5} \textrm{ Pa.s}$, $c_p=1.005\times 10^3  \textrm{ J.K.kg$^{-1}$}$,
and  $\kappa=2.5\times 10^{-2}\textrm{ W.m$^{-1}$.K$^{-1}$}$ respectively, while the atmospheric pressure is $P_0=1.013 \times 10^5 \textrm{ Pa}$. 
This lead to the bulk modulus $B_0=\gamma P_0$.

The material in the layer is supposed to be made of the periodic orthorhombic lattice of overlapping ellipsoids filled with air. 
As shown in Fig. \ref{fig:FIG2} (a), the rigid frame in its unit cell is obtained by extrusion of the ellipsoid having semi-axes 
$r_I=0.66\,\ell $, $r_{I\!\!I}=1.32\,\ell$, and $r_{I\!\!I\!\!I}=1.98\,\ell$ in the directions $(\e_{I},\e_{I\!\!I},\e_{I\!\!I\!\!I})$, from the parallelepiped cube having sizes 
$l_I=\ell$, $l_{I\!\!I}=2\,\ell$, and $l_{I\!\!I\!\!I}=3\,\ell$ in that same coordinate system.  
Under the condition of the scale separation $3 k_0 \ell \ll 1$, the theory of  two-scale asymptotic homogenization \cite{SanchezPalencia1980,Auriault2009} can be applied to the lattice to describe it as an effective fluid material satisfying the governing equations (\ref{eq:GoverningEqnsA}). Due to symmetries in the unit cell, the effective density tensor is diagonal in the Cartesian coordinate system $\mathcal{R}_\rho=(\e_{I},\e_{I\!\!I},\e_{I\!\!I\!\!I})$. Each principal density $\rho_J$ with   $J=I,I\!\!I,I\!\!I\!\!I$  of this diagonal density tensor, as well as the effective bulk modulus $B$ are then approximated according to the following Johnson--Champoux--Allard--Lafarge (JCAL) formulas \cite{Johnson1987,Lafarge1997}, 
 %----------------------------------------------------------------------------------%
\begin{subequations}
	\begin{align}
		%.................................................................................................................................................%
		&\frac{\rho_J}{\rho_0} 
		= \frac{ \alpha^\infty_J}{\phi} + \I \,  \frac{ \eta / K^0_J }{ \omega \rho_0  }\sqrt{1-\frac{\I  \omega\rho_0}{\eta}   \left(\frac{2\alpha^{\infty}_J  K^0_J}{\phi\Lambda_J}\right)^2} ,
		\label{eq:JCALa} \\
		%.................................................................................................................................................%
		&\frac{\gamma-1}{  \gamma - \displaystyle \frac{B_0}{ \phi B}  } 
		 =  1+ \I  \, \frac{  \phi \kappa /  \Theta^0 }{\omega  \rho_0 c_p  }\sqrt{  1-\frac{\I  \omega \rho_0 c_p  }{\kappa  }\left(\frac{2  \Theta^0 }{\phi\Lambda'}\right)^2}.
		\label{eq:JCALb} %
		%.................................................................................................................................................%
	\end{align}
\label{eq:JCAL}%
\end{subequations}
%----------------------------------------------------------------------------------%
Here, $\phi$ is the porosity,  $ K^0_J$, $\alpha^\infty_J$ and $\Lambda_J$ are the visco-static permeability, the high frequency tortuosity, and the characteristic viscous length in the direction $\e_{J}$ with $J=I,I\!\!I,I\!\!I\!\!I$, and $ \Theta^0$ and $\Lambda'$ are the thermo-static permeability and the  characteristic thermal length. All these parameters are defined from periodic cell problems provided by the theory of  two-scale asymptotic homogenization \cite{SanchezPalencia1980,Auriault2009}. These latter, recalled in the Supplementary Material, are solved numerically by means of the Finite Element Method using the software COMSOL Multiphysics\,\textregistered . The values of the calculated JCAL parameters are 
 %----------------------------------------------------------------------------------%
\begin{subequations}
	\begin{align}
		%.................................................................................................................................................%
		&			K^0_{I} =  0.11\, \ell^2,
		&&		K^0_{I\!\! I} = 0.08\,  \ell^2,
		&&		K^0_{I\!\!I\!\!I} =  0.09\,  \ell^2,
		\label{eq:JCALparamA} \\
		%.................................................................................................................................................%
		&			\alpha^\infty_{I} = 1.18 , 
		&&		\alpha^\infty_{I\!\! I} = 1.06 , 
		&&		\alpha^\infty_{I\!\!I\!\!I} = 1.04 , 
		\label{eq:JCALparamB} \\
		%.................................................................................................................................................%
		&			\Lambda_{I} = 1.02\,  \ell,
		&&		\Lambda_{I\!\! I} = 1.20\,  \ell,
		&&		\Lambda_{I\!\!I\!\!I} = 1.34\,  \ell,
		\label{eq:JCALparamC} \\
		%.................................................................................................................................................%
		&			\phi=0.91,
		&&		\Theta^0=0.20\,  \ell^2,
		&&		\Lambda'=1.84\,  \ell.
		\label{eq:JCALparamD} %
		%.................................................................................................................................................%
	\end{align}
\label{eq:JCALparam}%
\end{subequations}
%----------------------------------------------------------------------------------%
In what follows, the characteristic pore size $\ell=200\textrm{ $\mu$m}$ has been chosen so that $3 k_0\ell \approx 0.11$ at $ 10\textrm{ kHz}$,
which guarantees that the condition of scale separation is satisfied sharply in the frequency range $[10\textrm{ Hz}, 10\textrm{ kHz}]$ over which the retrieval method will be applied. 
With such a sharp separation of scales, the Drude boundary layers at the layer interfaces $\Gamma_0$ and $\Gamma_L$ can be neglected \cite{Levy1977,SanchezPalencia1980}. Moreover, the 
layer includes 150 unit cells in its thickness, which guarantees a bulk behaviour of the material. 

Now, the diagonal matrix density $\boldsymbol{\rho}^{\star} $ with principal densities given by Eq.~(\ref{eq:JCALa}) is rotated by the roll, pitch an yaw angles $\theta_{I}=\pi/6$, $\theta_{I\!\!I}=\pi/4$, and $\theta_{I\!\!I\!\!I}=\pi/3$ to result in the fully-anisotropic density tensor $\boldsymbol{\rho}= \mathbf{R}\cdot  \boldsymbol{\rho}^{\star} \cdot \,^t\mathbf{R}$ with $\mathbf{R}=\mathbf{R}_3\left(\theta_{I\!\!I\!\!I} \right)\mathbf{R}_2\left(\theta_{I\!\!I} \right) \mathbf{R}_1\left(\theta_{I} \right)$. Reflection and transmission coefficients related to incident plane waves impinging the layer are then computed 
by means of the Finite Element Method using the software COMSOL Multiphysics\,\textregistered,  thus ensuring that the inverse crime is not committed. 

To illustrate the generality of the retrieval method, different angles of incidence have been considered for the different tests, although they can very well be identical in practice. 
To form the pair of in-plane wavevectors $ \textbf{k}_\Gamma = \pm k'_1 \e_1$ having orientations of the incident plane given by the angles $\psi=0$ and $\psi=\pi$, the angle of incidence 
$\varphi=\pi/3$ has been used. To form the pair of in-plane wavevectors $ \textbf{k}_\Gamma = \pm k'_2 \e_2 $ in the planes of incidence oriented by $\psi=\pm\pi/2$ , the 
the angle of incidence  $\varphi=\pi/6$  has been used. Finally, the angle of incidence $\varphi=\pi/4$ in the plane of incidence oriented by $\psi=\pi/3$ has been considered to 
yield the in-plane wavevector $ \textbf{k}_\Gamma = k''_1 \e_1 + k_2'' \e_2$, see Sec.~\ref{sec:InverseProblem}. 

The retrieval method has been applied to give access to the bulk modulus and the six components of the density tensor. The reconstructed normalized bulk modulus $B^{\dag}/P_0$ and normalized components of the density tensor $\rho^{\dag}_{ij}/\rho_0$, $i,j, \in [1,2,3]$ are given in Fig. \ref{fig:FIG2}(b-c) and Fig. \ref{fig:FIG2}(d-e). The retrieved parameters are in excellent agreement with those used in the direct problem. However, the real and imaginary parts of the non-diagonal components do not have the expected signs to comply with causality. This apparent features is actually due to the fact that the reference coordinate system do not coincide with the principal directions of the material. To lift this ambiguity, the density tensor has been diagonalized and the normalized principal densities $\rho_I^{\dag}/\rho_0$, $\rho_{I\!\!I}^{\dag}/\rho_0$ and $\rho_{I\!\!I\!\!I}^{\dag}/\rho_0$ are given in Fig. \ref{fig:FIG2}(g-h). These principal densities comply with the causality conditions, $\textrm{Re}\left(\rho_J^{\dag}\right)\geq0$ and $\textrm{Im}\left(\rho_J^{\dag}\right)\geq0$, $J=I, I\!\!I, I\!\!I\!\!I$.
To sort the densities along the principal directions, an orthonormal basis is first built $(x_{I}^{\dag},x_{I\!\!I}^{\dag},x_{I\!\!I\!\!I}^{\dag})$ from the highest frequency reconstructed eigenvector matrix, which is compared with the provided principal directions Fig. \ref{fig:FIG2}(f). The eigenvectors are then compared (simple scalar product with the reconstructed vectors) to this basis to sort the densities. Note that the reconstructed principal directions are rotated when compared with the data but are correctly estimated. The additional recovery of the high frequency limit of the tortuosities $\alpha_J^{\infty}$, viscous characteristic lengths $\Lambda_J$, and static viscous permeabilities $K_J^0$ in the principal directions $J=I, I\!\!I, I\!\!I\!\!I$ and the open porosity $\phi$, thermal characteristic length $\Lambda'$, and static thermal permeability $\Theta^0$ is out of the scope of the present article but may be achieved by adapting existing methods such as Ref. \cite{Niskanen2017}.

%%%%%%%%%%%%%%%%%%%%%%%%%%%%%%%%%%%%%%%%%%%%%%%%%%%%%%%%%%%%%%%%%%
%%%%%%%%%%%%%%%%%%%%%%%%%%%%%%%%%%%%%%%%%%%%%%%%%%%%%%%%%%%%%%%%%%

%%%%%%%%%%%%%%%%%%%%%%%%%%%%%%%%%%%%%%%%%%%%%%%%%%%%%%%%%%%%%%%%%%
%%%%%%%%%%%%%%%%%%%%%%%%%%%%%%%%%%%%%%%%%%%%%%%%%%%%%%%%%%%%%%%%%%
\section{Conclusion}

Anisotropic fluids are of growing interest, mainly due to the rapid development of acoustic metamaterials, but also because many acoustic materials can be modeled as anisotropic fluids. A general method to characterize anisotropic fluid layers is developed and validated on simulated data in this work. This method extends existing ones based on the inversion of the scattering matrix to general three-dimensional anisotropic fluids whose principal directions are unknown and possibly tilted relative to the layer coordinate system. The method relies on the measurement of both transmission and reflection coefficients of the layer at a small number of incidence angles. From the transmission coefficients at two pairs of angles, the phase terms due to the possible out-of-plane principal directions are first recovered. Then, four pairs of transmission/reflection coefficients (possibly involving the same previous transmission coefficients) are required to recover analytically the six components of the symmetric density matrix and the bulk modulus. The density matrix is finally diagonalized to estimate the principal directions and the densities along the principal directions. This procedure is successfully applied to recover the principal directions, density matrix and bulk modulus of a simulated anisotropic viscothermal fluid material made of the orthorhombic lattice of overlapping ellipsoids. This procedure paves the way for the characterization and design of three-dimensional anisotropic metamaterials and acoustic materials.

%%%%%%%%%%%%%%%%%%%%%%%%%%%%%%%%%%%%%%%%%%%%%%%%%%%%%%%%%%%%%%%%%%
%%%%%%%%%%%%%%%%%%%%%%%%%%%%%%%%%%%%%%%%%%%%%%%%%%%%%%%%%%%%%%%%%%

%%%%%%%%%%%%%%%%%%%%%%%%%%%%%%%%%%%%%%%%%%%%%%%%%%%%%%%%%%%%%%%%%%
%%%%%%%%%%%%%%%%%%%%%**** ACKNOWLEDGMENTS ****%%%%%%%%%%%%%%%%%%%%%%%%%%%%
%%%%%%%%%%%%%%%%%%%%%%%%%%%%%%%%%%%%%%%%%%%%%%%%%%%%%%%%%%%%%%%%%%

\begin{acknowledgments}
The authors gratefully acknowledge ANR \textit{Chaire industrielle} MACIA (ANR-16-CHIN-0002) %
and RFI Le Mans Acoustique (R\'egion Pays de la Loire) PavNat project. %
This article is based upon work from COST Action DENORMS CA15125, supported by COST (European Cooperation in Science and Technology).
\end{acknowledgments}
%%%%%%%%%%%%%%%%%%%%%%%%%%%%%%%%%%%%%%%%%%%%%%%%%%%%%%%%%%%%%%%%%%
%%%%%%%%%%%%%%%%%%%%%%%%%%%%%%%%%%%%%%%%%%%%%%%%%%%%%%%%%%%%%%%%%%

%%%%%%%%%%%%%%%%%%%%%%%%%%%%%%%%%%%%%%%%%%%%%%%%%%%%%%%%%%%%%%%%%%
%%%%%%%%%%%%%%%%%%%%%**** BIBLIOGRAPHY ****%%%%%%%%%%%%%%%%%%%%%%%%%%%%%%%
%%%%%%%%%%%%%%%%%%%%%%%%%%%%%%%%%%%%%%%%%%%%%%%%%%%%%%%%%%%%%%%%%%
%\bibliographystyle{jasanum}
%\bibliography{BibtexAnisoCharact}
%%%%%%%%%%%%%%%%%%%%%%%%%%%%%%%%%%%%%%%%%%%%%%%%%%%%%%%%%%%%%%%%%%
%%%%%%%%%%%%%%%%%%%%%%%%%%%%%%%%%%%%%%%%%%%%%%%%%%%%%%%%%%%%%%%%%%
%

%%%%%%%%%%%%%%%%%%%%%%%%%%%%%%%%%%%%%%%%%%%%%%%%%%%%%%%%%%%%%%%%%%
%%%%%%%%%%%%%%%%%%%%%%% ****** End of file ******%%%%%%%%%%%%%%%%%%%%%%%%%%%%%%
%%%%%%%%%%%%%%%%%%%%%%%%%%%%%%%%%%%%%%%%%%%%%%%%%%%%%%%%%%%%%%%%%%
\end{document}

\preprint{}
%Title
	\title{Supplementary material to:\\ General method to retrieve all effective acoustic properties of fully-anisotropic fluid materials in three dimensional space}
	%General acoustic characterization of three-dimensional homogeneous anisotropic fluid layers
%%%%%%%%%%%%%%%%%%%%%%%%%%%%%%%%%%%%%%%%%%%%%%%%%%%%%%%%%%%%%%%%%%
%Authors and affiliation
	\author{Arthur~Terroir}
		%\email{Arthur.Terroir.Etu@univ-lemans.fr}
	\author{Logan~Schwan}
	%\email{Logan.Schwan@univ-lemans.fr}
	\author{Th\'eo~Cavalieri}
		%\email{Theo.Cavalieri@univ-lemans.fr}
	\author{Vicente~Romero-Garc\'ia}
		%\email{Vicente.Romero@univ-lemans.fr}
	\author{Gw\'ena\"el~Gabard}
		%\email{Gwenael.Gabard@univ-lemans.fr}
	\author{Jean-Philippe~Groby}
		\email{Jean-Philippe.Groby@univ-lemans.fr}
	\affiliation{Laboratoire d'Acoustique de l'Universit\'e du Mans, LAUM - UMR CNRS 6613, Le Mans Universit\'e, Avenue Olivier Messiaen, 72085 LE MANS CEDEX 9, France}
%Date
	\date{\today}
%%%%%%%%%%%%%%%%%%%%%%%%%%%%%%%%%%%%%%%%%%%%%%%%%%%%%%%%%%%%%%%%%%
\begin{abstract}
Here, the supplementary material to the paper entitled ``General method to retrieve all effective acoustic properties of fully-anisotropic fluid materials in three dimensional space" is presented. 
In Sec.~\ref{sec:ApparentProperties}, we derive the equations of apparent mass and momentum conservation in the normal direction of the  anisotropic layer from those involving the physical fields in the three-dimensional space. 
In Sec.~\ref{sec:Nicolson}, the Nicolson--Ross--Weir formula for anisotropic media are derived in details. 
In Sec.~\ref{sec:DensityTensor},  the analytical expressions of the coefficients of the density tensor are provided as functions of parameters assessed during the retrieval method. 
In Sec.~\ref{sec:CellProblems}, the cell problems leading to the Johnson--Champoux--Allard--Lafarge parameters in the homogneization of the orthorombic lattice of overlapping ellipsoids are recalled.
\end{abstract}
%%%%%%%%%%%%%%%%%%%%%%%%%%%%%%%%%%%%%%%%%%%%%%%%%%%%%%%%%%%%%%%%%%
%Make title page
	\maketitle
%%%%%%%%%%%%%%%%%%%%%%%%%%%%%%%%%%%%%%%%%%%%%%%%%%%%%%%%%%%%%%%%%%

%%%%%%%%%%%%%%%%%%%%%%%%%%%%%%%%%%%%%%%%%%%%%%%%%%%%%%%%%%%%%%%%%%
%%%%%%%%%%%%%%%%%%%%%%%%%%%%%%%%%%%%%%%%%%%%%%%%%%%%%%%%%%%%%%%%%%
\section{Apparent equations of mass and momentum conservation}
\label{sec:ApparentProperties}

We begin by recalling the  equations of mass and momentum conservation for the anisotropic material $\Omega$:  
%----------------------------------------------------------------------------------%
\begin{subequations}
	\begin{align}
		&  \textrm{i}\omega P\big/ B=  \nabla\cdot \mathbf{V}  ,
		\label{eq:GoverningEqnsA}\\
		& \textrm{i}\omega  \mathbf{V}= \boldsymbol{H} \nabla P ,
		\label{eq:GoverningEqnsB}%
	\end{align}
\label{eq:GoverningEqns}%
\end{subequations}
%----------------------------------------------------------------------------------%
where the symmetric tensor $\boldsymbol{H}=\boldsymbol{\rho}^{-1}$ is the inverse of the density tensor. 
We denote  $\boldsymbol{H}_\Gamma= H_{ij}$ with $(i,j)\in\{1,2\}^2$ its restriction to in-plane directions, and 
$\textbf{q}= q_1\e_1+q_2\e_2$ the vector with components: 
%----------------------------------------------------------------------------------%
\begin{equation}
	q_1 =    H_{13}/H_{33},
\qquad\text{and}\qquad
	q_2 =    H_{23}/H_{33}. 
\label{eq:ApparentParameters}
\end{equation}
%----------------------------------------------------------------------------------%
Furthermore, the pressure and velocity fields are taken in the following form prescribed by the incident plane wave,  the linearity in the system, and the homogeneity of the layer: 
%----------------------------------------------------------------------------------%
\begin{equation}
	P=\widehat{P}(x_3)e^{\textrm{i}k_1 x_1+\textrm{i}k_2 x_2} \ \ \text{and} \ \  \mathbf{V}=\widehat{\mathbf{V}}(x_3) e^{\textrm{i}k_1 x_1+\textrm{i}k_2 x_2}, 
\label{eq:FieldsInLayer}
\end{equation}
%----------------------------------------------------------------------------------%
where $\widehat{P}(x_3)$ and $\widehat{\mathbf{V}}(x_3) $ still depend on the coordinate $x_3$. 
The in-plane vector is defined by $\textbf{k}_\Gamma = k_1 \e_1 + k_2 \e_2 $. The in-plane and normal components of the velocity $\widehat{\mathbf{V}}(x_3) $ are also identified as 
%----------------------------------------------------------------------------------%
\begin{equation}
	\widehat{\mathbf{V}} = \widehat{\mathbf{V}}_\Gamma + \widehat{V}_3\e_3, 
\label{eq:VelocityComponents}
\end{equation}
%----------------------------------------------------------------------------------%
where $\widehat{\mathbf{V}}_\Gamma = \widehat{V}_1\e_1 + \widehat{V}_2\e_2,$ and $\widehat{V}_j = \widehat{\mathbf{V}} \cdot \e_j$ for $j\in\{1,2,3\}$.
Using the fact that derivation of the fields $P$ and $\mathbf{V}$ in Eq.~(\ref{eq:FieldsInLayer})  with respect to $x_j$  is equivalent to their multiplication by $\textrm{i} k_j$ for $j\in\{1,2\}$,  the conservation equations in Eq.~(\ref{eq:GoverningEqns}) can be re-written as follows, where Eq.~(\ref{eq:GoverningEqns2A}) is equivalent to Eq.~(\ref{eq:GoverningEqnsA}) while 
Eqs.~(\ref{eq:GoverningEqns2B}) and (\ref{eq:GoverningEqns2C}) are equivalent to Eq.~(\ref{eq:GoverningEqnsB}) in the in-plane and normal directions respectively:
%----------------------------------------------------------------------------------%
\begin{subequations}
	\begin{align}
		&  \frac{\textrm{i}\omega  \widehat{P}}{ B }=   \textrm{i}  \textbf{k}_\Gamma \cdot  \widehat{\mathbf{V}}_\Gamma    + \frac{\partial  \widehat{V}_3}{\partial x_3},
		\label{eq:GoverningEqns2A}\\
		& \textrm{i}\omega  \mathbf{V}_\Gamma =   \textrm{i}\boldsymbol{H}_\Gamma .\textbf{k}_\Gamma \widehat{P} + H_{33} \frac{\partial \widehat{P} }{\partial x_3}  \textbf{q},
		\label{eq:GoverningEqns2B} \\
		& \textrm{i}\omega  \widehat{V}_3  =  H_{33}  \(   \textrm{i} \textbf{q}\cdot \textbf{k}_\Gamma \widehat{P} +   \frac{\partial \widehat{P} }{\partial x_3} \).
		\label{eq:GoverningEqns2C}%
	\end{align}
\label{eq:GoverningEqns2}%
\end{subequations}
%----------------------------------------------------------------------------------%
Equation (\ref{eq:GoverningEqns2C}) yields the apparent momentum conservation in the normal direction, and involves the apparent density $\widetilde{\rho}\, $: 
%----------------------------------------------------------------------------------%
\begin{equation}
	 \textrm{i} \omega \widetilde{\rho}\,   \widehat{V}_3 =  \textrm{i} (\textbf{q}\cdot \textbf{k}_\Gamma )  \widehat{P} + \frac{ \partial \widehat{P}}{\partial x_3}
\ \ \text{where}\ \
	\widetilde{\rho}\,   = \frac{1}{H_{33}}.
\label{eq:ApparentMomentumConservation}
\end{equation}
%----------------------------------------------------------------------------------%
Besides, substitution of Eq.~(\ref{eq:GoverningEqns2B}) in (\ref{eq:GoverningEqns2A}) provides the relation: 
%----------------------------------------------------------------------------------%
\begin{equation}
  \frac{\textrm{i}\omega  \widehat{P}}{ B }=   
 \textrm{i} \textbf{k}_\Gamma \cdot \textbf{q} \frac{ H_{33}}{\textrm{i} \omega}  \frac{\partial \widehat{P} }{\partial x_3}  
- \frac{  \widehat{P} }{\textrm{i} \omega}   \textbf{k}_\Gamma \cdot (\boldsymbol{H}_\Gamma .\textbf{k}_\Gamma)
+ \frac{\partial  \widehat{V}_3}{\partial x_3},
\label{eq:ApparentMassConservation0}
\end{equation}
%----------------------------------------------------------------------------------%
while, according to Eq.~(\ref{eq:GoverningEqns2C}), the $x_3$-derivative of $\widehat{P}$ reads: 
%----------------------------------------------------------------------------------%
\begin{equation}
	\frac{\partial \widehat{P} }{\partial x_3} =  \frac{ \textrm{i}\omega }{H_{33}}  \widehat{V}_3 -  \textrm{i} \textbf{q}\cdot \textbf{k}_\Gamma \widehat{P} .
\label{eq:DerivativeP}
\end{equation}
%----------------------------------------------------------------------------------%
Substitution of Eq.~(\ref{eq:DerivativeP}) in (\ref{eq:ApparentMassConservation0}) yields the apparent equation of mass conservation:
%----------------------------------------------------------------------------------%
\begin{equation}
	\frac{\textrm{i} \omega \widehat{P}}{ \widetilde{B} } =    \textrm{i} (\textbf{q}\cdot \textbf{k}_\Gamma )  \widehat{V}_3  + \frac{\partial \widehat{V}_3 }{ \partial x_3 }  , 
\label{eq:ApparentMassConservation}
\end{equation}
%----------------------------------------------------------------------------------%
where the apparent bulk modulus $ \widetilde{B}$ satisfies:
%----------------------------------------------------------------------------------%
\begin{equation}
	\frac{1}{ \widetilde{B} } =   \frac{1}{ B }  -   \frac{ 1 }{ \omega^2  }   \(  \:^t\textbf{k}_\Gamma . \boldsymbol{H}_\Gamma .\textbf{k}_\Gamma -    H_{33}  (\textbf{k}_\Gamma \cdot \textbf{q} )^2    \)  .
\label{eq:ApparentBulkModulus}
\end{equation}
%----------------------------------------------------------------------------------%
Expansion of the right-hand side of Eq.~(\ref{eq:ApparentBulkModulus}) gives:
%----------------------------------------------------------------------------------%
\begin{equation}
\begin{array}{l}
	\displaystyle 
		 \:^t\textbf{k}_\Gamma . \boldsymbol{H}_\Gamma .\textbf{k}_\Gamma -    H_{33}  (\textbf{k}_\Gamma \cdot \textbf{q} )^2 
				= H_{11} k_1^2 + H_{22} k_2^2  \\
	\displaystyle 
			\quad + 2H_{12} k_1 k_2- H_{33}\( q_1^2 k_1^2 + 2 q_1q_2k_1k_2 + q_2^2 k_2^2 \).
\end{array}
\label{eq:Expand}
\end{equation}
%----------------------------------------------------------------------------------%
Substitution of  Eq.~(\ref{eq:Expand}) into (\ref{eq:ApparentBulkModulus}) yields
%----------------------------------------------------------------------------------%
\begin{subequations}
	\begin{align}
		&\frac{\omega^2}{B} -\frac{\omega^2}{ \widetilde{B}(k_1,k_2)  }   = \xi_{11}  k_1^2 +  \xi_{22}k_2^2   +2 \,  \xi_{12} k_1k_2 , 
		\label{eq:ApparentBulk2A}\\
		&\text{with}\quad  \xi_{ij} = H_{ij}  - H_{33} \,  q_i \, q_j ,\ \  \forall (i,j)\in\{1,2\}^2. 
		\label{eq:ApparentBulk2B}%
	\end{align}
\label{eq:ApparentBulk2}%
\end{subequations}
%----------------------------------------------------------------------------------%
Equations (\ref{eq:ApparentMomentumConservation}), (\ref{eq:ApparentMassConservation}) and (\ref{eq:ApparentBulk2}) provide the results of this section.

%%%%%%%%%%%%%%%%%%%%%%%%%%%%%%%%%%%%%%%%%%%%%%%%%%%%%%%%%%%%%%%%%%
%%%%%%%%%%%%%%%%%%%%%%%%%%%%%%%%%%%%%%%%%%%%%%%%%%%%%%%%%%%%%%%%%%

%%%%%%%%%%%%%%%%%%%%%%%%%%%%%%%%%%%%%%%%%%%%%%%%%%%%%%%%%%%%%%%%%%
%%%%%%%%%%%%%%%%%%%%%%%%%%%%%%%%%%%%%%%%%%%%%%%%%%%%%%%%%%%%%%%%%%
\section{Nicolson--Ross--Weir formula in anisotropic media}
\label{sec:Nicolson}

To derive the Nicolson--Ross--Weir formula in anisotropic media, we start with the following system stating the boundary conditions at the interfaces of the layer with the surrounding medium:
%----------------------------------------------------------------------------------%
\begin{equation}
	\begin{Bmatrix} 	R+1\\	 \\		\displaystyle \frac{R-1}{\widetilde{Z}_0} \\			\end{Bmatrix}
= \textbf{U} \cdot
	\begin{bmatrix} 	e^{ \I  k_3^{-}L} & 0\\  	\\	0 & e^{\I  k_3^{+} L}\\		\end{bmatrix}
	\cdot 
	\textbf{U}^{-1}
	\cdot
	\begin{Bmatrix} 	T \\	\\		\displaystyle	 \frac{-T }{\widetilde{Z}_0} \\			\end{Bmatrix},
\label{eq:Sol2}
\end{equation}
%----------------------------------------------------------------------------------%
where matrices $\textbf{U}$ and $\textbf{U}^{-1}$ are given by
%----------------------------------------------------------------------------------%
\begin{equation}
 \textbf{U} = 
	\begin{bmatrix}
		    \widetilde{Z}  & \widetilde{Z} \\
		    -1 & 1 \\
	\end{bmatrix}, \quad \text{and }\quad 
 \textbf{U}^{-1} = \frac{1}{2}
	\begin{bmatrix}
	     1\big/\widetilde{Z}  & -1\\
		1\big/\widetilde{Z}  & 1 \\
	\end{bmatrix},
\label{eq:Sol}%
\end{equation}
%----------------------------------------------------------------------------------%
while wavenumbers in the layer read as follows in the $x_3$-direction: 
%----------------------------------------------------------------------------------%
\begin{equation}
	k_3^{\pm} = -\textbf{q}\cdot \textbf{k}_\Gamma \pm\widetilde{k}.
\label{eq:LayerWavenumber}
\end{equation}
%----------------------------------------------------------------------------------%
Substitution of Eq.~(\ref{eq:Sol}) into (\ref{eq:Sol2}) yields the following two equations once developed: 
%----------------------------------------------------------------------------------%
\begin{subequations}
	\begin{align}
	%................................................................................................................%
	&& 2(R+1) ~~    &=   T e^{ \I  k_3^{+}L} (1-\sigma)  + T e^{ \I  k_3^{-}L} (1+\sigma),
	\label{eq:TheTwoEqnsA}\\
	%................................................................................................................%	
	&& 2(R+1) \sigma  &= T e^{ \I  k_3^{+}L}  (1-\sigma)  -T e^{ \I  k_3^{-}L} (1+\sigma)   ,
	\label{eq:TheTwoEqnsB}%
	%................................................................................................................%
	\end{align}
\label{eq:TheTwoEqns}%
\end{subequations}
%----------------------------------------------------------------------------------%
where $\sigma=\widetilde{Z}/\widetilde{Z}_0$ is the ratio of apparent impedances. Summing side by side Eqs.~(\ref{eq:TheTwoEqnsA}) and (\ref{eq:TheTwoEqnsB}) while using the expression (\ref{eq:LayerWavenumber}) of the wavenumber $ k_3^{+}$ provides: 
%----------------------------------------------------------------------------------%
\begin{equation}
	( T  e^{- \I  \textbf{q}\cdot \textbf{k}_\Gamma L  } ) e^{+ \I \widetilde{k}L} = 1 + \frac{1+\sigma}{1-\sigma} R.
\label{eq:SumTwoEqns}%
\end{equation}
%----------------------------------------------------------------------------------%
Subtracting side by side Eq.~(\ref{eq:TheTwoEqnsB}) to (\ref{eq:TheTwoEqnsA}) provides 
%----------------------------------------------------------------------------------%
\begin{equation}
	( T  e^{- \I  \textbf{q}\cdot \textbf{k}_\Gamma L  } ) e^{- \I \widetilde{k}L} = 1 + \frac{1-\sigma}{1+\sigma} R.
\label{eq:DifTwoEqns}%
\end{equation}
%----------------------------------------------------------------------------------%
Multiplying side by side Eqs.~(\ref{eq:SumTwoEqns}) and (\ref{eq:DifTwoEqns}) leads to: 
%----------------------------------------------------------------------------------%
\begin{equation}
	( T  e^{- \I  \textbf{q}\cdot \textbf{k}_\Gamma L  } )^2  = \(  1 + \frac{1-\sigma}{1+\sigma} R \)\(1 + \frac{1+\sigma}{1-\sigma} R \).
\label{eq:MultwoEqns}%
\end{equation}
%----------------------------------------------------------------------------------%
Equation (\ref{eq:MultwoEqns}) can be re-written in the form 
%----------------------------------------------------------------------------------%
\begin{equation}
	\frac{1+\sigma^2}{1-\sigma^2} = \frac{1+R^2-( T  e^{- \I  \textbf{q}\cdot \textbf{k}_\Gamma L  } )^2}{2R}.
\label{eq:MultwoEqns2}%
\end{equation}
%----------------------------------------------------------------------------------%
Solving for $\sigma^2$ in Eq.~(\ref{eq:MultwoEqns2}) yields
%----------------------------------------------------------------------------------%
\begin{equation}
	\sigma^2  = \frac{1+R^2-( T  e^{- \I  \textbf{q}\cdot \textbf{k}_\Gamma L  } )^2  -2R }{1+R^2-( T  e^{- \I  \textbf{q}\cdot \textbf{k}_\Gamma L  } )^2  +2R}.
\label{eq:Sigma}%
\end{equation}
%----------------------------------------------------------------------------------%
Taking the square root of Eq.~(\ref{eq:Sigma}), the apparent impedance is found: 
%----------------------------------------------------------------------------------%
\begin{equation}
	\widetilde{Z}=\pm \widetilde{Z}_0
		\sqrt{ \frac{(1-R)^2-( T  e^{- \I  \textbf{q}\cdot \textbf{k}_\Gamma L  } )^2  }{(1+R)^2-( T  e^{- \I  \textbf{q}\cdot \textbf{k}_\Gamma L  } )^2}}, 
\label{eq:MyZZ}%
\end{equation}
%----------------------------------------------------------------------------------%
while Eqs.~(\ref{eq:TheTwoEqnsA}) and (\ref{eq:TheTwoEqnsB})  can be cast in the single relation
%----------------------------------------------------------------------------------%
\begin{equation}
	 e^{ \mp \I \widetilde{k}L} =  \( 1 + \frac{1 \mp\widetilde{Z}/\widetilde{Z}_0}{ 1 \pm\widetilde{Z}/\widetilde{Z}_0 } R \) \frac{1}{T  e^{- \I  \textbf{q}\cdot \textbf{k}_\Gamma L  }}.
\label{eq:TwoEqnsSingle}%
\end{equation}
%----------------------------------------------------------------------------------%
Equations (\ref{eq:MyZZ}) and (\ref{eq:TwoEqnsSingle})  provide the results of this section.

%%%%%%%%%%%%%%%%%%%%%%%%%%%%%%%%%%%%%%%%%%%%%%%%%%%%%%%%%%%%%%%%%%
%%%%%%%%%%%%%%%%%%%%%%%%%%%%%%%%%%%%%%%%%%%%%%%%%%%%%%%%%%%%%%%%%%

%%%%%%%%%%%%%%%%%%%%%%%%%%%%%%%%%%%%%%%%%%%%%%%%%%%%%%%%%%%%%%%%%%
%%%%%%%%%%%%%%%%%%%%%%%%%%%%%%%%%%%%%%%%%%%%%%%%%%%%%%%%%%%%%%%%%%
\section{Coefficients of the density tensor}
\label{sec:DensityTensor}

Here, we derive the expression of the six coefficients of the density tensor $ \boldsymbol{\rho} $ from the knowledge of the following parameters: 
%----------------------------------------------------------------------------------%
\begin{subequations}
	\begin{align}
		&\	q_1 =    H_{13}/H_{33}, \quad  q_2 =    H_{23}/H_{33},\quad  \widetilde{\rho}\,  =1/H_{33} ,  
		\label{eq:InitialParametersA}\\
		&\xi_{ij} = H_{ij}  - H_{33} \,  q_i \, q_j ,\ \  \forall (i,j)\in\{1,2\}^2. 
		\label{eq:InitialParametersB}%
	\end{align}
\label{eq:InitialParameters}%
\end{subequations}
%----------------------------------------------------------------------------------%
Since $ \boldsymbol{H} =  \boldsymbol{\rho}^{-1}$, tensors $\boldsymbol{H}$ and $\boldsymbol{\rho}$ satisfy
%----------------------------------------------------------------------------------%
\begin{equation}
	\boldsymbol{H} = \frac{ \:^t\text{comat}(\boldsymbol{\rho})}{\det(\boldsymbol{\rho})},
\quad\text{and}\quad
	\boldsymbol{\rho} = \det(\boldsymbol{\rho}) \:^t\text{comat}(\boldsymbol{H}),
\label{eq:MutualInverse}
\end{equation}
%----------------------------------------------------------------------------------%
where $\text{comat}(\cdot)$ stands for the comatrix. Since  $ \boldsymbol{H} $ and $ \boldsymbol{\rho}$ are symmetric, calculation of their co-matrices and of the determinant of the density tensor provide:
%----------------------------------------------------------------------------------%
\begin{widetext}
\begin{equation}
	\boldsymbol{H} = \frac{1}{\det(\boldsymbol{\rho})}
		\begin{bmatrix}
			%.....................................................................................................%
			\quad&			\rho_{22} \rho_{33} 	- 	\rho_{23}^2 
			&\qquad& 		\rho_{13} \rho_{23}	-	\rho_{33}\rho_{12} 
			&\quad& 		\rho_{12} \rho_{23}	-	\rho_{22}\rho_{13} 
			& \quad
			\\  \\
			%.....................................................................................................%
			\quad&				\rho_{13} \rho_{23} 	- 	\rho_{33} \rho_{12}
			&\qquad& 		\rho_{11} \rho_{33} 	- 	\rho_{13}^2
			&\qquad& 		\rho_{12}\rho_{13}	-	\rho_{11}\rho_{23}
			& \quad
			\\ \\
			%.....................................................................................................%
			\quad&			\rho_{12}\rho_{23} 	- 	\rho_{13}\rho_{22}
			&\qquad&		\rho_{12}\rho_{13}	- 	\rho_{11}\rho_{23}
			&\qquad&		\rho_{11}\rho_{22} 	- 	\rho_{12}^2
			& \quad
			\\
			%.....................................................................................................%
		\end{bmatrix},
\label{eq:Comat}
\end{equation}
\begin{equation}
	\boldsymbol{\rho} =  \det(\boldsymbol{\rho}) \cdot
		\begin{bmatrix}
			%.....................................................................................................%
			\quad&			H_{22} H_{33} 	- 	H_{23}^2 
			&\qquad& 		H_{13} H_{23}	-	H_{33}H_{12} 
			&\quad& 		H_{12} H_{23}	-	H_{22}H_{13} 
			& \quad
			\\  \\
			%.....................................................................................................%
			\quad&				H_{13} H_{23} 	- 	H_{33} H_{12}
			&\qquad& 		H_{11} H_{33} 	- 	H_{13}^2
			&\qquad& 		H_{12}H_{13}	-	H_{11}H_{23}
			& \quad
			\\ \\
			%.....................................................................................................%
			\quad&			H_{12}H_{23} 	- 	H_{13}H_{22}
			&\qquad&		H_{12}H_{13}	- 	H_{11}H_{23}
			&\qquad&		H_{11}H_{22} 	- 	H_{12}^2
			& \quad
			\\
			%.....................................................................................................%
		\end{bmatrix},
\label{eq:ComatH}
\end{equation}
\begin{equation}
	\text{det}(\boldsymbol{\rho}) = 
		\rho_{13}\,( \rho_{12}\rho_{23}-\rho_{13}\rho_{22} ) + \rho_{23}\,(\rho_{12}\rho_{13}-\rho_{11}\rho_{23}) + \rho_{33}\,( \rho_{11} \rho_{22}-\rho_{12}^2).
\label{eq:Deter}
\end{equation}
\end{widetext}
%----------------------------------------------------------------------------------%
In what follows, the following coefficient is used: 
%----------------------------------------------------------------------------------%
\begin{equation}
	\Delta_{12} = \rho_{11} \rho_{22}- \rho_{12}^2 = H_{33} \det(\boldsymbol{\rho})
\label{eq:Delta12}
\end{equation}
%----------------------------------------------------------------------------------%
Combining Eqs.~(\ref{eq:InitialParametersA}) to (\ref{eq:Delta12}), the following relations are found immediately: 
%----------------------------------------------------------------------------------%
\begin{subequations}
	\begin{align}
		& 	q_1  =   \frac{H_{13}}{H_{33}} = \frac{ \rho_{12} \rho_{23}	-	\rho_{22} \rho_{13} }{\Delta_{12}},
		\label{eq:FirstRelationsA} \\
		& 	q_2  =   \frac{H_{23}}{H_{33}} = \frac{ \rho_{12} \rho_{13}	-	 \rho_{11} \rho_{23} }{\Delta_{12}},
		\label{eq:FirstRelationsB} \\
		& \det(\boldsymbol{\rho}) = \Delta_{12} \(  \rho_{13} q_1 + \rho_{23} q_2 + \rho_{33} \), 
		\label{eq:FirstRelationsC} \\
		&  \widetilde{\rho}\,  =\frac{1}{H_{33} } = \frac{\det(\boldsymbol{\rho})}{ \Delta_{12}} =   \rho_{13} q_1 + \rho_{23} q_2 + \rho_{33} .
		\label{eq:FirstRelationsD} 
	\end{align}
\label{eq:FirstRelations}%
\end{subequations}
%----------------------------------------------------------------------------------%
Now, the parameters $\xi_{ij}$ in Eq.~(\ref{eq:InitialParametersB}) can be re-written
%----------------------------------------------------------------------------------%
\begin{subequations}
	\begin{align}
		&&  \xi_{11} = H_{11}  - H_{33} \,  q_1^2     &= \frac{H_{11}H_{33}-H_{13}^2}{H_{33}}  , \\
		&&  \xi_{22} = H_{22}  - H_{33} \,  q_2^2     &= \frac{H_{22}H_{33}-H_{23}^2}{H_{33}}  , \\
		&&  \xi_{12} = H_{12}  - H_{33} \,  q_1q_2   &= \frac{H_{12}H_{33}-H_{13}H_{23}}{H_{33}}  .
	\end{align}
\label{eq:Xi1}%
\end{subequations}
%----------------------------------------------------------------------------------%
Besides, equation   (\ref{eq:ComatH})   provides the following coefficients of the density tensor: 
%----------------------------------------------------------------------------------%
\begin{subequations}
	\begin{align}
		& \rho_{11} =  \( H_{22}H_{33}-H_{23}^2 \) \text{det}(\boldsymbol{\rho}), \\
		& \rho_{22} =  \( H_{11}H_{33}-H_{13}^2 \) \text{det}(\boldsymbol{\rho}), \\
		& \rho_{12} =  \( H_{13}H_{23}-H_{33}H_{12} \) \text{det}(\boldsymbol{\rho}) .
	\end{align}
\label{eq:Xi2}%
\end{subequations}
%----------------------------------------------------------------------------------%
Combination of Eqs.~(\ref{eq:Delta12}), (\ref{eq:Xi1}) and (\ref{eq:Xi2}) yields
%----------------------------------------------------------------------------------%
\begin{equation}
	 \xi_{11} = 	\frac{ \rho_{22} }{\Delta_{12}} ,
\quad
	 \xi_{22} =  \frac{\rho_{11}}{ \Delta_{12}} , 
\quad
	\xi_{12} = \frac{ - \,\rho_{12} }{\Delta_{12}} . 
\label{eq:Xi3}%
\end{equation}
%----------------------------------------------------------------------------------%
These parameters obviously satisfy the equality
%----------------------------------------------------------------------------------%
\begin{equation}
	\xi_{11}  \xi_{22} -  \xi_{12}^2 = \frac{1}{\Delta_{12}} = \frac{1}{\rho_{11}\rho_{22}-\rho_{12}^2}.
\label{eq:Xi4}%
\end{equation}
%----------------------------------------------------------------------------------%
As a result, Eq.~(\ref{eq:Xi3}) can be inverted to give
%----------------------------------------------------------------------------------%
\begin{subequations}
	\begin{align}
		&\rho_{11}  = \  \ \xi_{22} \Big/(  \xi_{11}  \xi_{22} -  \xi_{12}^2  ) ,\\
		& \rho_{22} = \ \   \xi_{11}  \Big/(  \xi_{11}  \xi_{22} -  \xi_{12}^2   ) ,\\
		&\rho_{12}=  - \,  \xi_{12}   \Big/(  \xi_{11}  \xi_{22} -  \xi_{12}^2) . 
	\end{align}
\label{eq:RhoInPlane}%
\end{subequations}
%----------------------------------------------------------------------------------%
Now combining Eqs.~(\ref{eq:FirstRelations}a,b) and (\ref{eq:Xi3}), the parameters $q_1$ and $q_2$ read: 
%----------------------------------------------------------------------------------%
\begin{equation}
		q_1 = -( \xi_{11} \rho_{13} + \xi_{12} \rho_{23}  ) ,
\quad
		q_2 = -( \xi_{22} \rho_{23} + \xi_{12} \rho_{13}  ) . 
\label{eq:q1q2}%
\end{equation}
%----------------------------------------------------------------------------------%
Solving for $\rho_{13}$ and $\rho_{23}$ in Eq.~(\ref{eq:q1q2}) yields: 
%----------------------------------------------------------------------------------%
\begin{subequations}
	\begin{align}
		&\rho_{13}  = (\xi_{12} q_2 - \xi_{22} q_1)  \Big/(  \xi_{11}  \xi_{22} -  \xi_{12}^2  ) ,\\
		& \rho_{23} = (\xi_{12} q_1 - \xi_{11} q_2)   \Big/(  \xi_{11}  \xi_{22} -  \xi_{12}^2   ) .
	\end{align}
\label{eq:RhoCoupling}%
\end{subequations}
%----------------------------------------------------------------------------------%
Finally, substitution of Eq.~(\ref{eq:RhoCoupling}a,b) into (\ref{eq:FirstRelationsD}) leads to:
%----------------------------------------------------------------------------------%
\begin{equation}
	\rho_{33} =  \widetilde{\rho}\,  + (   \xi_{22} q_1^2   +\xi_{11} q_2^2     -2\, \xi_{12} q_1 q_2   )   \Big/(  \xi_{11}  \xi_{22} -  \xi_{12}^2  ).
\label{eq:Rho33}%
\end{equation}
%----------------------------------------------------------------------------------%
Equations (\ref{eq:RhoInPlane}a,b,c), (\ref{eq:RhoCoupling}a,b) and (\ref{eq:Rho33}) provide the result of this section.

%%%%%%%%%%%%%%%%%%%%%%%%%%%%%%%%%%%%%%%%%%%%%%%%%%%%%%%%%%%%%%%%%%
%%%%%%%%%%%%%%%%%%%%%%%%%%%%%%%%%%%%%%%%%%%%%%%%%%%%%%%%%%%%%%%%%%
\section{Cell problems in the homogenization process}
\label{sec:CellProblems}

In this section, the definition of the Johnson--Champoux--Allard--Lafarge \cite{Johnson1987,Lafarge1997} parameters are provided in terms of volume and surface averages of periodic potential fields over the unit cell of the porous medium. These potential fields satisfy periodic cell problems, the full derivation of which from the theory of two-scale asymptotic homogenization \cite{SanchezPalencia1980,Auriault2009} is not recalled here. 
In what follows, $\Omega_\uc$ denotes the unit cell of the porous medium, $\Omega_\ff$ denotes the fluid domain inside the unit cell, and $\Gamma$ denotes the fluid/solid interface with $\vect{n}$ as normal vector. Moreover, the volume average of a field  defined in the fluid domain of  the unit cell is denoted by 
%----------------------------------------------------------------------------------%
\begin{equation}
	\langle \cdot \rangle = \frac{1}{| \Omega_\uc|} \int_{\Omega_\ff}{ \cdot \, \text{d}\Omega}, 
\end{equation}
%----------------------------------------------------------------------------------%
where $| \Omega_\uc| $ is the overall volume of the unit cell. In the present case of the parallelepiped cube with sizes $l_I$, $l_{I\!\!I}$, and $l_{I\!\!I\!\!I}$ in the directions $(\e_{I},\e_{I\!\!I},\e_{I\!\!I\!\!I})$, the volume of the unit cell is $| \Omega_\uc| = l_I \, l_{I\!\!I} \, l_{I\!\!I\!\!I}$. Furthermore, the porosity of the porous medium reads 
%----------------------------------------------------------------------------------%
\begin{equation}
	\phi= \langle 1 \rangle = | \Omega_\ff | \big/ | \Omega_\uc|.
\end{equation}
%----------------------------------------------------------------------------------%

\subsection{Visco-inertial cell problem}

The frequency-dependent visco-inertial cell problem consists of the incompressible Stokes flow through the unit cell in response to a unit pressure gradient applied in the $\e_{J}$  with   $J=I,I\!\!I,I\!\!I\!\!I$: 
%-------------------------------------------------------------------------------%
\begin{equation}
\left\lbrace
	\begin{array}{l}
		\displaystyle 
			\text{div}( \vect{k}_J ) = 0,  \\ 
		\displaystyle 
			   \text{div}(   \textbf{grad}(\vect{k}_J) )  +  \frac{\I \omega \rho_0}{ \eta}  \vect{k}_J =  \textbf{grad}(   \zeta_J )  - \vect{e}_J  ,  \\ 
		\displaystyle 
			\vect{k}_J  \equiv \vect{0}\text{\quad on }\Gamma, \\
		\displaystyle 
			 \langle  \zeta_J \rangle \equiv 0, \\
		\displaystyle 
			\vect{k}_J,\quad \zeta_J \quad  \Omega\text{-periodic},\\
	\end{array}
\right.
\label{eq:ViscoInertialCellPb}
\end{equation}
%-------------------------------------------------------------------------------%
where $\vect{k}_J$ plays the role of the velocity field and $\zeta_J$ of its associated pressure field. The visco-inertial frequency-dependent permeability tensor reads: 
%----------------------------------------------------------------------------------%
\begin{equation}
	\tens{K} =  \langle    \vect{e}_J  \otimes \vect{k}_J  \rangle,
\end{equation}
%----------------------------------------------------------------------------------%
with implicit summation on $J$, and with $\otimes $ being the tensor product.  Due to symmetries in the unit cell, directions $(\e_{I},\e_{I\!\!I},\e_{I\!\!I\!\!I})$ are identified as principal axes of the visco-inertial permeability tensor, which is therefore diagonal in this coordinate system  $\tens{K} = \textbf{diag}(K_J)$ where $K_J=\langle  \vect{k}_J  \cdot  \vect{e}_J \rangle$. 

The visco-static permeability tensor corresponds to  the tensor $\tens{K}$ at the frequency $\omega=0$. The fields $(\vect{k}_J,\zeta_J)=(\vect{k}^0_J,\zeta^0_J)$ at $\omega=0$ satisfy the geometric cell problem: 
%-------------------------------------------------------------------------------%
\begin{equation}
\left\lbrace
	\begin{array}{l}
		\displaystyle 
			\text{div}( \vect{k}_J^0 ) = 0,  \\ 
		\displaystyle 
			   \text{div}(   \textbf{grad}(\vect{k}_J^0) )  =  \textbf{grad}(   \zeta_J ^0)  - \vect{e}_J^0  ,  \\ 
		\displaystyle 
			\vect{k}_J^0  \equiv \vect{0}\text{\quad on }\Gamma, \\
		\displaystyle 
			 \langle  \zeta_J^0 \rangle \equiv 0, \\
		\displaystyle 
			\vect{k}_J^0,\quad \zeta_J^0 \quad  \Omega\text{-periodic}, \\
	\end{array}
\right.
\label{eq:ViscoInertialCellPb0}
\end{equation}
%-------------------------------------------------------------------------------%
and the visco-static permeability $K_J^0 $ is real and positive valued and is given by: 
%----------------------------------------------------------------------------------%
\begin{equation}
	K_J^0 =  \langle  \vect{k}_J^0  \cdot  \vect{e}_J \rangle.
\label{eq:Kj0}
\end{equation}
%----------------------------------------------------------------------------------%
At high frequencies $\omega\rightarrow\infty$, the fields are denoted  $(\vect{k}_J,\zeta_J)=(\vect{k}^\infty_J,\zeta^\infty_J)$ and satisfy the cell problem: 
%-------------------------------------------------------------------------------%
\begin{equation}
\left\lbrace
	\begin{array}{l}
		\displaystyle 
			\text{div}( \vect{k}_J ^\infty) = 0,  \\ 
		\displaystyle 
			   \frac{\I \omega \rho_0}{ \eta}  \vect{k}_J^\infty =  \textbf{grad}(   \zeta_J^\infty )  - \vect{e}_J  ,  \\ 
		\displaystyle 
			\vect{k}_J^\infty  \cdot \vect{n} = \vect{0}\text{\quad on }\Gamma, \\
		\displaystyle 
			 \langle  \zeta_J^\infty \rangle \equiv 0, \\
		\displaystyle 
			\vect{k}_J^\infty,\quad \zeta_J^\infty \quad  \Omega\text{-periodic}.\\
	\end{array}
\right.
\label{eq:ViscoInertialCellPbInfty}
\end{equation}
%-------------------------------------------------------------------------------%
The high-frequency limit of the principal permeability  then reads: 
%----------------------------------------------------------------------------------%
\begin{equation}
	K_J^\infty =  \frac{\phi \eta}{-\I \omega \rho_0 \alpha_J^\infty}
\quad\text{where}\quad
	\alpha_J^\infty = \frac{\phi}{\phi     -  \displaystyle  \left\langle  \frac{ \partial  \zeta_J^\infty}{\partial x_J } \right\rangle} 	, 
\label{eq:KjInf}
\end{equation}
%----------------------------------------------------------------------------------%
with $\alpha_J^\infty$ being  the high-frequency tortuosity. Besides, the characteristic viscous length $\Lambda_J$  is related to the high-frequency asymptotic limit of the permeability $K_J$ when the viscous boundary layer at the fluid/solid interface is accounted for \cite{Johnson1987}. It reads: 
%----------------------------------------------------------------------------------%
\begin{equation}
	\Lambda_J = 2 \frac{  \int_{\Omega_\ff}{ \vect{k}_J^\infty \cdot \vect{k}_J^\infty  \,\text{d}\Omega}}{  \int_{\Gamma}{ \vect{k}_J^\infty \cdot \vect{k}_J^\infty  \,\text{d}\Gamma} }.
\label{eq:Lvis}
\end{equation}
%----------------------------------------------------------------------------------%
Equations (\ref{eq:ViscoInertialCellPb0}) and (\ref{eq:ViscoInertialCellPbInfty}) provide the cell problems solved to obtain the JCAL parameters $K_J^0$, $\alpha_J^\infty$ and $\Lambda_J $ given in Eqs. (\ref{eq:Kj0}), (\ref{eq:KjInf}), and  (\ref{eq:Lvis}).

\subsection{Thermo-acoustic cell problem}

The frequency-dependent thermo-acoustic cell problem consists of the heat conduction through the unit cell in response to a constant pressure applied:
%-------------------------------------------------------------------------------%
\begin{equation}
\left\lbrace
	\begin{array}{l}
		\displaystyle 
			   \text{div}(   \textbf{grad}(\vartheta) )  +  \frac{\I \omega \rho_0 c_p }{ \kappa }  \vartheta = -1  ,  \\ 
		\displaystyle 
			 \vartheta   \equiv 0 \text{\quad on }\Gamma, \\
		\displaystyle 
			\vartheta \quad  \Omega\text{-periodic},\\
	\end{array}
\right.
\label{eq:ThermocCellPb}
\end{equation}
%-------------------------------------------------------------------------------%
where $\vartheta$ plays the role of the temperature field. The thermo-acoustic frequency-dependent permeability tensor reads: 
%----------------------------------------------------------------------------------%
\begin{equation}
	\Theta =  \langle \vartheta   \rangle.
\end{equation}
%----------------------------------------------------------------------------------%
The thermostatic permeability  corresponds to  the permeability $\Theta$ at the frequency $\omega=0$. The field  $ \vartheta= \vartheta^0$ at $\omega=0$ satisfy the geometric cell problem: 
%-------------------------------------------------------------------------------%
\begin{equation}
\left\lbrace
	\begin{array}{l}
		\displaystyle 
			   \text{div}(   \textbf{grad}(\vartheta^0) )   = -1  ,  \\ 
		\displaystyle 
			 \vartheta^0   \equiv 0 \text{\quad on } \Gamma, \\
		\displaystyle 
			\vartheta^0 \quad  \Omega\text{-periodic}.\\
	\end{array}
\right.
\label{eq:ThermocCellPb0}
\end{equation}
%-------------------------------------------------------------------------------%
and the thermostatic permeability $\Theta^0 $ is given by: 
%----------------------------------------------------------------------------------%
\begin{equation}
	\Theta^0 =  \langle \vartheta^0  \rangle. 
\label{eq:Theta0}
\end{equation}
%----------------------------------------------------------------------------------%
At high frequencies $\omega\rightarrow\infty$, the thermal field is denoted  $\vartheta= \vartheta^\infty $ and is uniform over the cell. It satisfies:
%----------------------------------------------------------------------------------%
\begin{equation}
 \frac{\I \omega \rho_0 c_p }{ \kappa }  \vartheta^\infty = -1  
\quad\text{and}\quad
	\Theta^\infty =  \langle \vartheta^\infty  \rangle = \frac{\phi  \kappa }{-\I \omega \rho_0 c_p } . 
\label{eq:ThetaInf}
\end{equation}
%----------------------------------------------------------------------------------%
Besides, the characteristic thermal length $\Lambda'$  is related to the high-frequency asymptotic limit of the permeability $\Theta$ when the thermal boundary layer at the fluid/solid interface is accounted for. It reads: 
%----------------------------------------------------------------------------------%
\begin{equation}
	\Lambda' = 2 \frac{  \int_{\Omega_\ff}{  \vartheta^\infty \cdot  \vartheta^\infty  \,\text{d}\Omega}}{  \int_{\Gamma}{   \vartheta^\infty \cdot  \vartheta^\infty \,\text{d}\Gamma} }
	= 2 \frac{|\Omega_\ff|}{| \Gamma|}.
\label{eq:Lthe}
\end{equation}
%----------------------------------------------------------------------------------%
Equation (\ref{eq:ThermocCellPb0})  provides the cell problem  solved to obtain the JCAL parameters $\Theta^0$    and $\Lambda' $ given in Eqs. (\ref{eq:Theta0})  and  (\ref{eq:Lthe}). 

\pagebreak

%%%%%%%%%%%%%%%%%%%%%%%%%%%%%%%%%%%%%%%%%%%%%%%%%%%%%%%%%%%%%%%%%%
%%%%%%%%%%%%%%%%%%%%%%%%%%%%%%%%%%%%%%%%%%%%%%%%%%%%%%%%%%%%%%%%%%

%
%
%%%%%%%%%%%%%%%%%%%%%%%%%%%%%%%%%%%%%%%%%%%%%%%%%%%%%%%%%%%%%%%%%%%
%%%%%%%%%%%%%%%%%%%%%%**** ACKNOWLEDGMENTS ****%%%%%%%%%%%%%%%%%%%%%%%%%%%%
%%%%%%%%%%%%%%%%%%%%%%%%%%%%%%%%%%%%%%%%%%%%%%%%%%%%%%%%%%%%%%%%%%%
%
%\begin{acknowledgments}
%The authors gratefully acknowledge the support from the ANR \textit{Chaire industrielle} MACIA (ANR-16-CHIN-0002) %
%and from the RFI Le Mans Acoustique (R\'egion Pays de la Loire) PavNat project. %
%This article is based upon work from COST Action DENORMS CA15125, supported by COST (European Cooperation in Science and Technology).
%\end{acknowledgments}
%%%%%%%%%%%%%%%%%%%%%%%%%%%%%%%%%%%%%%%%%%%%%%%%%%%%%%%%%%%%%%%%%%%
%%%%%%%%%%%%%%%%%%%%%%%%%%%%%%%%%%%%%%%%%%%%%%%%%%%%%%%%%%%%%%%%%%%
%

%%%%%%%%%%%%%%%%%%%%%%%%%%%%%%%%%%%%%%%%%%%%%%%%%%%%%%%%%%%%%%%%%%
%%%%%%%%%%%%%%%%%%%%%**** BIBLIOGRAPHY ****%%%%%%%%%%%%%%%%%%%%%%%%%%%%%%%
%%%%%%%%%%%%%%%%%%%%%%%%%%%%%%%%%%%%%%%%%%%%%%%%%%%%%%%%%%%%%%%%%%
%\bibliographystyle{jasanum}
%\bibliography{BibtexAnisoCharact}
%%%%%%%%%%%%%%%%%%%%%%%%%%%%%%%%%%%%%%%%%%%%%%%%%%%%%%%%%%%%%%%%%%
%%%%%%%%%%%%%%%%%%%%%%%%%%%%%%%%%%%%%%%%%%%%%%%%%%%%%%%%%%%%%%%%%%

%

%%%%%%%%%%%%%%%%%%%%%%%%%%%%%%%%%%%%%%%%%%%%%%%%%%%%%%%%%%%%%%%%%%
%%%%%%%%%%%%%%%%%%%%%%% ****** End of file ******%%%%%%%%%%%%%%%%%%%%%%%%%%%%%%
%%%%%%%%%%%%%%%%%%%%%%%%%%%%%%%%%%%%%%%%%%%%%%%%%%%%%%%%%%%%%%%%%%